%% file: disk.tex
\documentstyle[aaspp4]{article}
\def\etal{\emph{et~al.\ }}
\def\qmin{$Q_{\rm min}$}
\def\mrat{$M_D/M_*$}

\begin{document}
\title{Dynamics of Circumstellar Disks}

\author{Andrew F. Nelson}
\affil{Department of Physics, The University of Arizona, Tucson AZ 85721}
\author{Willy Benz}
\affil{Steward Observatory, The University of Arizona, Tucson AZ 85721\ and  \\
Physikalishes Institut, Universitaet Bern, Sidlerstrasse 5, Ch-3012 Bern,
Switzerland}
\author{Fred C. Adams}
\affil{Department of Physics, The University of Michigan, Ann Arbor MI 48109}
\author{David Arnett}
\affil{Steward Observatory, The University of Arizona, Tucson AZ 85721}

\begin{abstract}
We present a series of 2-dimensional hydrodynamic simulations of
massive disks around protostars. We simulate the same physical problem 
using both a `Piecewise Parabolic Method' (PPM) code and a `Smoothed 
Particle Hydrodynamic' (SPH) code, and analyze their differences.

The disks studied here range in mass from $0.05 M_*$ to 
$1.0 M_*$ and in initial minimum Toomre $Q$ value from $1.1$ to
$3.0$. We adopt simple power laws for the initial density and
temperature in the disk with an isothermal ($\gamma=1$) equation
of state. The disks are locally isothermal. We allow the central
star to move freely in response to growing perturbations. The
simulations using each code are compared to discover differences
due to error in the methods used. For this problem, the strengths of
the codes overlap only in a limited fashion, but similarities exist
in their predictions, including spiral arm pattern speeds and
morphological features. Our results represent limiting cases (i.e. 
systems evolved isothermally) rather than true physical systems.  

Disks become active from the inner regions outward. From the earliest
times, their evolution is a strongly dynamic process rather than a 
smooth progression toward eventual nonlinear behavior. Processes that
occur in both the extreme inner and outer radial regions affect the growth 
of instabilities over the entire disk. Effects important for the global
morphology of the system can originate at quite small distances from the 
star. We calculate approximate growth rates for the spiral patterns;
the one-armed ($m=1$) spiral arm is not the fastest growing 
pattern of most disks. Nonetheless, it plays a significant role due
to factors which can excite it more quickly than other patterns.
A marked change in the character of spiral structure occurs with varying
disk mass. Low mass disks form filimentary spiral structures with many
arms while high mass disks form grand design spiral structures with few
arms.

In our SPH simulations, disks with initial minimum $Q=1.5$ or lower 
break up into proto-binary or proto-planetary clumps.  However, these 
simulations cannot follow the physics important for the flow and must
be terminated before the system has completely evolved. At their 
termination, PPM simulations with similar initial conditions show uneven
mass distributions within spiral arms, suggesting that clumping behavior
might result if they were carried further. Simulations of tori, for which
SPH and PPM are directly comparable, do show clumping in both codes. 
Concern that the point-like nature of SPH exaggerates clumping, that our 
representation of the gravitational potential in PPM is too coarse, and
that our physics assumptions are too simple, suggest caution in 
interpretation of the clumping in both the disk and torus simulations. 
\end{abstract}


\doublespace

\section{Introduction}\label{intro}
Over the past several years a broad paradigm of star formation has
been developed (see Shu, Adams \&
Lizano 1987). First, a cloud of gas and dust
collapses and forms a protostar with a surrounding disk.  Later
the star/disk system ejects matter in outflows as well as
continuing to accrete matter from the cloud.  Finally, accretion and
outflow cease and the star gradually loses its disk and evolves onto
the main sequence. While this paradigm provides for a good qualitative
picture of the star formation process, many important issues 
require further work. For example, observations by several groups
(Simon \etal 1995, Ghez \etal 1993, Leinert \etal 1993, Reipurth \&
Zinnecker 1993) show that young stars in many different star forming
regions are commonly found in binary or higher order multiple systems,
with a broad peak in separation distance at around 30 AU.  In
addition, many of the higher order multiples show hierarchical
characteristics: a distant companion orbiting a close binary for
example. In what manner are multiple systems such as these formed?

A variety of studies have been undertaken to model the processes
leading to the observed systems. One class of models begins with the
collapse of a cloud of matter. These results (Bate \etal 1995, Foster
\& Boss 1996, Boss 1995, Burkert \& Bodenheimer 1993, Bonnell \&
Bastien 1992, Myhill \& Kaula 1992) show that both single stars and
multiple systems can be formed from the collapse and subsequent
fragmentation of rotating, spherical or elongated molecular cloud
cores.  This class of simulations focus on the collapse phase but do
not follow in detail the dynamics of disks formed from the material with
initially higher angular momentum.

In addition, a number of models extended beyond the initial collapse
(Bonnell 1994, Pickett \etal 1996, Woodward \etal 1994) have shown
that post-collapse objects can be driven into fragmentation, or into
spiral arm and bar formation prior to the development of a Keplerian 
disk.  Laughlin \& Bodenheimer (1994) have simulated the evolution of a
collapsing cloud in 2D and then followed its late time behavior with a
3D disk simulation. They have found that such a collapse leads to a
core plus a long lived, broad torus. The development of $m=1$ and
$m=2$ spiral patterns may lead to late time fragmentation of the torus
($m$ is the number of arms in the spiral pattern).

As a star-disk or multiple-star-disk system evolves, the dynamics of
the disk itself as well as its interaction with the star or binary becomes
important in determining the final configuration of the system. Depending
on its mass and temperature, a disk may develop spiral density waves 
and viscous phenomena of varying importance. Each may be capable of
processing matter through the disk as well as influencing how the
disk eventually decays away as the star evolves onto the main sequence.

A variety of mechanisms for production of spiral instabilities in
disks around single stars have been suggested. An incomplete list 
includes the linear perturbation results of Adams, Ruden \& Shu 
(1989) (hereafter ARS) who suggest a mechanism (`SLING'-- see Shu
\etal 1990) by which a resonance between the 
star and a one-armed ($m$=1) spiral mode may become globally unstable.  
Both perturbation theory (Papaloizou \& Lin 1989) and numerical
calculation (Papaloizou \& Savonjie 1991, Heemskirk \etal 1992) have
shown another instability mechanism based on the distribution of
specific vorticity (termed ``vortensity'') which can influence
evolution in disks and tori. It is driven primarily by wave
interactions at corotation and can act either to suppress or amplify
spiral waves in the disk, depending on the vortensity gradient there.
Another family of instabilities is based upon vortensity
gradients at the boundaries of the disk or torus. The SWING amplifier 
(Goldreich \& Lynden-Bell 1965, Julian \& Toomre 1966, Goldreich \&
Tremaine 1978) provides an instability channel whereby low
amplitude leading spiral arms unwind and are transformed
into much larger amplitude trailing waves. A feedback cycle then 
creates additional leading waves and the instability grows. 

This paper is a continuation of work by two of us (Adams \& Benz 1992,
hereafter AB92), who began modeling of disks 
of mass $M_D\gtrsim 0.5 M_*$ and observed formation of
spiral arms and clumps. We present a series of two dimensional numerical
simulations of circumstellar disks
with masses between $0.05 M_*$ and $1.0 M_*$. We 
attempt to characterize the growth of instabilities and pay
particular attention to the existence and effect of the SLING 
instability.  In section \ref{codes}, we outline the 
numerical methods used and discuss the limitations of each code
and their effects on our simulations. In section \ref{phyasmpt}, we
outline the initial conditions adopted for the disks studied and in
section \ref{results}, we first describe qualitatively the results of
our simulations and then begin a quantitative analysis
of the pattern growth, the correspondence between
two hydrodynamic codes, and the correspondence between linear analyses
and hydrodynamic simulations. In Section \ref{summary}, we
summarize the results and their significance in the evolution of stars
and star systems.

\section{The Codes}\label{codes}
\subsection{Solving the Hydrodynamic Equations}

In order to understand the properties of protostellar disks
we have adapted two complementary hydrodynamic codes to the task
of simulating such evolution: the Smoothed Particle Hydrodynamic
(SPH) method and the Piecewise Parabolic Method (PPM).  These codes
use very different techniques for solving the equations of
hydrodynamics, and it is hoped that, by the use of such widely different
techniques, numerical artifacts can be sorted out from true physical
evolution. Each code has unique features that allow the
simulation of these systems in some regimes not accessible to the other.

The SPH method (see reviews by Benz 1990, Monaghan 1992) uses a
procedure by which hydrodynamic quantities and their derivatives 
are calculated from an
interpolation technique over neighboring particles. The interpolation
kernel used in our simulations is the standard B-spline kernel with
compact support.  The smoothing length $h$ is varied over time in a manner
such that the number of neighbors is approximately conserved, subject
to the condition that a minimum value of $h\sim R_D/1700$ (where $R_D$ 
is the disk radius) is set to ensure time steps do not become too small.
A second order Runge-Kutta-Fehlberg integrator which includes time step
control is used to evolve the system in time. Being gridless, the
main advantage of the SPH method in our context lies in its ability to
follow structure formation anywhere in the disk without the
limitations associated with a prescribed grid. The two main disadvantages
of the SPH technique are (1) the inherent random noise level associated 
with the discrete representation of the fluid and 
(2) the high shear component of the dissipation connected with the 
mathematical formulation of the artificial viscosity.

We also have adapted the PROMETHEUS hydrodynamic code (Fryxell,
M\"uller \& Arnett 1989, 1991) to the problem of evolving disks
around protostars. PROMETHEUS is based on the `Piecewise Parabolic
Method' (PPM) of Colella \& Woodward (1984) in which a high order
polynomial interpolation is used to determine cell edge values used in
calculating a second order solution to the Riemann shock tube problem at 
each cell boundary.  The interpolation is modified in regions of sharp
discontinuities to track shocks and contact discontinuities more closely
and retain their sharpness, while a monotonizing condition smoothes out
unphysical oscillations. The solution to the one-dimensional Riemann 
problem is then used to calculate fluxes and advance the solution in
time.  This code was selected because of its low numerical dissipation
and its excellent resolution of discontinuities and shocks.

Both codes incorporate self-gravity using modified versions of the
binary tree described in Benz \etal (1990) which approximates the
gravity of groups of distant particles in a multipole expansion while
calculating interactions of nearby particles explicitly. Gravitational
forces due to neighbor particles are softened to avoid divergences as
particles pass near each other. Due to the organization of the grid,
the tree construction can be considerably simplified in the PPM
version by substituting a procedure by which adjacent grid cells 
(modeled as point masses for the purpose of the gravity calculation)
or groups of grid cells become progressively higher nodes in the
tree. Two simulations run at higher resolution (simulations {\it pch2}
and {\it pch6} in table \ref{ppm-tabl} below) implemented an FFT
based solution to Poisson's equation (Binney \& Tremaine
1987, pp. 96ff). Results for a disk simulation at identical resolution
showed that the tree and the FFT solutions gave identical
dynamical results, however the FFT version proved to be substantially
faster. The torus simulations of section \ref{sphppm}, which are 
more sensitive to resolution, are also more sensitive to the implementation 
of the Poisson solver. In these cases the simulations using the tree code 
gave slower pattern growth rates than simulations using the FFT.

It is important to make a distinction between the resolution of the
hydrodynamics and that of the representation of the gravitational
potential. Just as PPM is well adapted for discontinuities, SPH is
well adapted for gravitational clumping.
The density reconstruction procedure utilized by PPM
contains more structure
than is available from the $N$ grid point algorithm used here.
Better resolution of the gravitational potential may be
possible using densities defined at both the cell centers and at cell
interfaces. Further, this grid effectively implies a gravitational
softening which is about one cell in size. This algorithm uses only the
cell center information, and references below to grid resolution in PPM
simulations will imply this fact.

\subsection{Viscosity in the Codes}\label{visc-sec}

Because disk evolution is partially driven by viscosity
in the disk, we must look carefully at issues related
to numerical viscosity. Except for codes based on a
local solution of the Riemann shock problem such as PPM, most methods
require implementation of an artificial viscosity to enforce stability
and/or improve the shock treatment by the code. In this regard SPH is
no exception and our version of the code implements the standard form
discussed in Benz (1990).  We use bulk and
von~Neumann-Richtmyer (so called `$\bar\alpha$' and `$\beta$')
viscosities to simulate viscous pressures which are linear and
quadratic in the velocity divergence. We incorporate a switch
(see Balsara 1995) which acts to reduce substantially the large
undesirable shear component associated with the standard form.

The bulk component of the artificial viscosity $\bar\alpha$ in the SPH
code can be identified with a kinematic viscosity $\nu$ (see Murray
1994) using the relation
\begin{equation}
\nu = {{ \bar\alpha c_s h}\over 8},
\end{equation}
where $c_s$ is the sound speed and $h$ is the smoothing length of a
particle. It is possible to  relate the coefficient of bulk artificial
viscosity $\bar\alpha$ to the $\alpha$-parameter of the standard viscous
prescription of accretion disks.  We equate the artificial viscosity to the
Shakura \& Sunyaev (1973) viscosity (defined by $\nu = \alpha c_s H$ and $H$ 
is the disk scale height defined as the local sound speed over the angular 
rotation rate $c_s/\Omega$). Solving for $\alpha$ yields 
\begin{equation}   \label{sph-alpha}
\alpha = {{ f\bar\alpha h \Omega}\over {8c_s} }, 
\end{equation}
where $f$ is the shear reduction factor discussed in Benz 1990 suitably
averaged over particles and time. 
We caution the reader that the identification of the SPH form of the
viscosity is not necessarily equivalent to that of the Shakura and
Sunyaev form, especially because of the approximate manner in which 
the Balsara switch must be taken into account. We estimate equation 
[\ref{sph-alpha}] may be valid to a factor of a few but should not be
taken as exact.

For a nearly Keplerian disk with a temperature $T \propto r^{-1/2}$
and a roughly linear variation of the smoothing length, $h$, with radius,
we obtain $\alpha \propto r^{-1/4}$. Depending on the temperature 
constant describing each disk ($T_0$, see section \ref{init}), $\alpha$ 
is of order $\sim10^{-2}$.  Only at small radii ($r\lesssim 2$AU) and low 
disk mass (for which $T_0$ becomes correspondingly small for a specified
value of \qmin) does $\alpha$ rise to values in the range
$\alpha\sim0.1-1$.  These values of $\alpha$ imply that the viscous
time scale, $\tau= r^2/\nu$, remains significantly longer than the few
orbital time scales we simulate. For most of
the disk, the SPH viscosity is small enough not to affect the
evolution of the disk significantly.  The von~Neumann ($\beta$) term
in the viscosity does not mirror the alpha prescription as the bulk
term does. Derived from the assumption that the viscosity is
proportional to square of the velocity divergence, its effect is limited
to portions of the flow in which shocks occur.

The numerical viscosity inherent to the PPM code is 
difficult to quantify. The nonlinear nature of the Riemann solver
(with the associated PPM `switches' to sharpen discontinuities and
enforce monotonicity) renders an artificial viscosity term
unnecessary.  However, a small numerical viscosity still appears
in the code. Porter \& Woodward (1994) derive fits for numerical
dissipation proportional to the third and fourth powers of ${\delta
x}/\lambda$ where ${\delta x}$ is a cell dimension and $\lambda$ is
the wavelength of a disturbance. Thus, large scale disturbances like
the spiral arms will experience little
dissipation, but small scale motions will be damped more.

\section{Physical Assumptions and Constraints}\label{phyasmpt}

Because our simulations involve dimensionless quantities such as the 
disk/star mass ratio and the Toomre stability parameter $Q$, the physics
itself is scalable to systems of different size. 
We shall express all quantities in 
units with values typical of the early stages of protostellar evolution. 
These units are also comparable (for the most massive disk simulations) to 
the final dimensions of our own solar system.  The star mass will be
assumed $M_* = 0.5 M_\odot$ and the disk radius $R_D$ = 50 AU. Time
units are given in either years or the disk orbit period defined by
$T_D=2\pi\sqrt{R_D^3/GM_*}$ which, with the mass and
radius given above, is equal to 500 years.

\subsection{Circumstellar Disk Initial Conditions} \label{init}

The initial conditions for prototype low and high mass disks are 
summarized graphically in figures \ref{dinit-ppm} and \ref{dinit-sph} 
for our PPM and SPH simulations respectively.  In functional form,
the disk mass is initially distributed according to a density power law
\begin{equation}
\Sigma(r) = \Sigma_0 \left[ 1 + \left({r\over r_c}\right)^2\right]^
{-{p\over{2}}},   \label{denslaw}
\end{equation}
where the power law exponent $p$ is set to 3/2. 
As we shall discuss in the following section we found that our PPM
simulations implementing the initial density profile of eq. [\ref{denslaw}]
became violently dynamic and we could not simulate the evolution of 
the system.  Instead, we have chosen to remove 
matter completely at small radii in our PPM runs by adopting an initial density
law which ensures that little matter remains at small radii or
interacts with the boundary. This density law takes the form
\begin{equation}\label{denslaw-a}
\Sigma(r) = {\Sigma_0{\left\{[{1-e^{\left({{r-R_0}\over{R_c}}\right)^2}]} 
		 \over{r}\right\}^p}},
\end{equation}
where $R_0$ is set to the radius of the innermost boundary cell and
$R_c$ is set arbitrarily to 6 AU. With this choice, the surface
density is substantially reduced near the inner boundary while
retaining a nearly pure $r^{-3/2}$ distribution for radii greater than
about 10 AU. 

The temperature is given by similar power law as
\begin{equation}
T(r) = T_0\left[1 + \left({r\over r_c}\right)^2\right]^{-{q\over{2}}},
    \label{templaw}
\end{equation}
with the exponent $q$ set to 1/2.  The softening radius $r_c$ for
both power laws is set to $r_c=R_D/50$(=1~AU).  $\Sigma_0$ 
and $T_0$ are determined from the disk mass and a choice of the
minimum value over the disk of the Toomre stability parameter $Q$, 
defined as
\begin{equation} \label{Qdef}
Q = {{\kappa c_s}\over{ \pi G \Sigma}},
\end{equation}
where $\kappa$ is the local epicyclic frequency.  For an ideal gas with
an isothermal equation of state (see section \ref{eos-sec}), the sound 
speed is defined as
\begin{equation}
c_s^2 = {{kT}\over{\mu m_p}},
\end{equation}
where the mean molecular weight is $\mu$ and we assume the gas is
of solar composition.  

We choose the value of the temperature power law index based on
observed temperature profiles in T~Tauri disks (see Beckwith \etal 
1990; Adams \etal 1990).  The density power law is much less well
constrained, and our choice of $p=3/2$ is roughly consistent with
the infall collapse calculations of Cassen \& Moosman (1981). 
As an additional motivation, this choice of exponents
matches the one adopted by ARS and allows a direct comparison with
their work.  We assume that the disks are vertically thin so
that two dimensional ($r$,$\phi$) simulations are justified.  The
variables of interest (density, pressure, etc.) are taken to
be vertically integrated quantities.  Magnetic fields are 
neglected in our simulations.

These temperature and density laws produce a profile for the 
instability parameter $Q$ that is nearly flat over the largest
portion of the disk, with a steep rise at small radii and a 
shallow increase towards the outer edge of the disk.  The minimum 
value of $Q$ in the disk is therefore located at $\sim 30-40$ AU,
depending upon the mass and temperature of a specific disk.
The $X$ parameter, important for SWING amplification and is defined by
\begin{equation}
X = {{r\kappa^2}\over{2\pi mG\Sigma}},
\end{equation}
with $m$ the number of spiral arms (the azimuthal wave number).  The $X$
parameter shows a similar pattern to that seen for $Q$, but with a steeper 
increase at large radii. In order for a system to be unstable to SWING 
amplification, the value of $X$ must be $\lesssim$3 in the region
of interest.  For most of the disks we study, $X$ is
larger than that required to keep the disk stable for the lowest
order spiral modes, so that we expect SWING not to contribute
to the growth of instabilities. Like the $Q$ and $X$ profiles, the 
vortensity profile shows a steep increase at small radii. In this 
case, such an increase may serve to stimulate growth due to 
the family of instabilities discussed by Papaloizou and Lin (1989).
We will discuss this possibility in more detail below.

The star is represented as a point mass, free to move in response
to gravitational forces from the surrounding disk. Initially, disk matter 
is placed on circular orbits around the star, with 
rotational equilibrium in the disk and radial velocities set to 
zero. Gravitational and pressure forces are balanced with centrifugal 
forces such that the rotation curve is given by
\begin{equation}
\Omega^2(r) = { {GM_*\over{r^3}} + {1\over{r}}{
       {\partial\Psi_D}\over{\partial{r}}} + {1\over{r}}{
       {{\bf\nabla}{P}}\over{\Sigma} } },    \label{rotlaw}
\end{equation}
where the symbols have their usual meanings and $\Psi_D$, the gravitational
potential of the disk, is calculated numerically with the same potential
solver utilized in the full hydrodynamic code.  The magnitudes 
of the pressure and gravitational forces are small compared to the stellar
term, therefore the disk is nearly Keplerian in character.

\subsection{Boundary Conditions, Construction of Circumstellar
Accretion Disks and Numerical Resolution} \label{bound}

To complete the specification of the initial state of
the systems, we must define the conditions at the boundaries 
of each simulation. The results of ARS suggest that the dynamics of
an accretion disk will be relatively insensitive to the implementation
of the inner boundary condition, becoming active only at distances far
from the star. On the other hand, the shape of outer edge of the disk
is predicted to be critical for the eventual growth of the SLING 
instability. In order to search for evidence of the SLING instability
we shall implement boundary conditions which may be favorable to its 
growth.

To ease time step constraints, we set the inner boundary at a greater
distance than that which is physically the case for a star/disk boundary.
With a grid code, we can define the inner boundary by modeling the inner
regions in some steady state approximation or by modifying the density 
law at small radii (in effect modeling tori) to reduce interactions with
the inner boundary. Since ARS predict that the inner regions of the disk
will be relatively stable, instabilities are not expected to grow there,
given a disk initially in rotational equilibrium. Any boundary condition
which does not perturb this equilibrium should be sufficient to describe
the inner disk.  Since by assumption, the inner disk begins in rotational
equilibrium (i.e. with $v_r=0$), no matter will cross the boundary and a 
simple reflecting boundary condition will suffice. The reflecting boundary
will also serve a second function. The four wave cycle
(Shu \etal 1990, hence STAR) important for the amplification of SLING 
requires a corotation or $Q$-barrier from which waves can be reflected
or refracted during part of the cycle. Until such resonances may develop 
on their own further out in the disk, the reflecting boundary serves as 
a surrogate for the actual resonances. 

Our PPM simulations showed that for a pure 
power law for the density (omitting the core radius of eq. 
[\ref{denslaw}]), the inner regions of the disk to be quite dynamic and
unstable. After a few orbits, matter in the inner
disk moved off its initial circular orbit and began interacting with
the boundary. The effect of these interactions is to give a
``kick'' to the system center of mass as matter reflects off the boundary.
In the worst cases, serious computational problems occurred after 
20-50 orbits of the inner disk edge and the calculations had to be 
stopped.

Several prescriptions for avoiding this behavior were attempted without
real success. These prescriptions included allowing matter to
accrete through the boundary onto the star, attaching the inner disk
matter to the star itself, treating the inner disk as a softened point 
mass at the origin with varying degrees of softening or by treating the 
inner disk matter as an additional point mass free to move in response 
to the star and the rest of the disk. In each case results obtained were
strongly dependent on the prescription followed. We conclude
that the dynamics important for the global
behavior of the physical system extend to quite small radii.

With this degree of activity in the inner disk it becomes reasonable
to assume that a portion of the inner disk matter becomes depleted by
accretion onto the star or ejected in an outflow on short time 
scales. The inner disk may expand in the $z$
direction and become truly three dimensional as the dynamical effects
create dissipation and heating. In light of these ideas,
and in order to concentrate our efforts on the large scale features,
we have chosen to implement the density law of eq. [\ref{denslaw-a}]
and study a system for which little mass exists close to the star
but which retains a power law profile further out. Due to the already
artificial nature of mass distribution at small radii, little physical
meaning can be attached to mass accretion rates through the inner 
boundary, therefore for simplicity we implement reflecting boundary
conditions to complete the specification of the inner grid edge.

For our SPH simulations, we define the inner boundary by establishing 
an accretion radius at a distance from the current position of the star 
of $R_D$/110(=0.4~AU). This distance is set to be slightly smaller
than the initial position of the innermost ring of particles in the
disk. The gravitational softening radius for the star is set to the same
value. As a particle's trajectory takes it inside the 
accretion/softening radius, its mass and momentum are added to the star
and it is removed from the calculation. This inner boundary 
condition does not prove to be as difficult to manage as in our PPM
simulations. Even though a great deal of activity occurs in the 
inner portion of the disk, no particular computational difficulties were 
experienced. We believe this activity is largely due to crude boundary 
conditions which obscure the true physical behavior of the system.
Particles near the boundary have no neighbor particles further
inward to provide pressure support, while accretion of a particle 
through the boundary implies a sudden loss of pressure support to 
its neighbors further out. Also, the stellar gravitational softening
reduces the effect of the star on the orbit of each SPH particle 
there. A small number particles near the boundary
are strongly affected.

Because of our interest in characterizing disk instabilities,
especially SLING, we have experimented with several outer boundary
conditions as well. In the PPM simulations we have implemented both 
a reflecting boundary and a boundary condition  in which matter falls 
onto the outer edge of the disk (an ``infall'' boundary). With the pure
reflecting conditions, we imitate the boundary conditions implemented
by ARS which have been identified as critical for the SLING instability. 
With the infall boundary condition, we relax this assumption slightly to 
allow the disk edge to begin outward expansion or begin to break up if
conditions require.

With the infall boundary, the outer disk edge is defined to be 
initially located at a cell interface several cells inward from the
outermost computational cell. We define the disk boundary assuming 
an isothermal shock, so that the density and radial component of the 
velocity are determined directly from the shock jump conditions. Since by
definition a shock implies that the tangential velocity across the 
shock is continuous, we know that at the disk edge, $R_D$, the 
$\phi$ component of the infall velocity is the same as the orbit 
velocity, $R_D\Omega(R_D)$. If we then specify the temperature
of the infalling gas as $T=10$ K, conservation laws for mass, momentum
and energy determine the flow from the shock to the outer grid radius. 
The infall is kept constant throughout the simulation at the 
values which initially define it. We note in passing that a flow
constructed in this manner is quite artificial and may have little
relation to flows in real systems. For our simulations,
infall provides a mechanism by which the outer edge of 
the disk can be reasonably well defined.

In our SPH simulations, we adopt a free outer boundary. This choice
has the advantage of simplicity in implementation, but suffers because
quantities such as the density or pressure are less well defined within
about two smoothing lengths of the boundary (see fig \ref{dinit-sph}).
The result is that over time the surface density at the disk edge spreads
radially to a width of $\sim$5 AU. The disk edge is no longer defined by
a sharp discontinuity, but does remain distinct except for very high
\qmin~simulations, for which the mass at the outer disk
edge is nearly unbound. The sharp outer boundary condition required
for SLING to become active is satisfied under these conditions.

At time zero in our SPH simulations we set approximately 8000 equal
mass particles on a series of concentric rings with the innermost ring
at a radius of $R_D$/100. For our PPM simulations we use a inner to
outer radius ratio of 50 and several grid resolutions. Our main series
of simulations, with reflecting outer boundary conditions, have a 
$64\times102$ cell cylindrical polar ($r,\phi$) grid. Two higher
resolution simulations are performed with a $100\times152$ grid, and
we have explored the use of an infall boundary at two resolutions
of $44\times64$ and $64\times96$. Grid cells are defined to be `squares'
in the sense that $\delta r = Cr\delta\phi$ over the entire grid, with
$C$ a constant $\sim$1. With the resolution used for our simulations,
SPH particle smoothing lengths are less than a few tenths of one AU
in the inner portion of the disk up to $\sim$1 AU in the outer disk.
Grid resolution in the PPM simulations is of order 0.1 AU at the inner
grid edge and $\sim$2 AU at the outer edge.

The relatively low resolution of our simulations results in part
from the large radial extent of the systems we study. Many of the 
important dynamical processes in a disk occur on time scales for 
orbits in the outer regions of the disk, but the size of a time step
(the Courant condition) is derived from the cell size at the inner
grid edge, where the cells are the smallest and the velocities are
largest.  Assuming nearly Keplerian rotation around the star, an 
inner grid radius at 1 AU, and a moderate resolution of order 150
azimuth cells, the time step is a few days, while the dynamical
time scale of the disk is a few $T_D=$500 yr. In order to evolve a
given simulation to completion, we must integrate over a half a 
million or more time steps. For `high' resolution simulations of say, 
300 or more azimuthal cells, the number is correspondingly increased. 
With the workstations available, it is not computationally feasible 
to run a large number of models to explore the relevant parameter space.
A similar problem exists for our SPH simulations.

\subsection{The Equation of State and Energy Considerations}\label{eos-sec}

In each code, a vertically integrated gas pressure is implemented
using a single component, `isothermal' ($\gamma=1$) gas equation of
state given by
\begin{equation}\label{eos}
P=\Sigma c_s^2.
\end{equation}
In PROMETHEUS (our version of the PPM
algorithm), a truly isothermal equation of state with $\gamma = 1$ is not 
easily obtained, therefore we use an `almost isothermal' ideal gas with
$\gamma = 1.01$ for these simulations. 

Each simulation is evolved isothermally, by which we mean that the temperature
of each cell, once defined at time zero (by eq. [\ref{templaw}] and an input 
value of \qmin), is fixed thereafter. Loss processes such as radiative cooling 
are assumed to balance local heating processes in the disk. Under this 
assumption, a packet of matter which moves radially inward or outward, 
heats or cools according to the prescribed temperature law. Matter which 
expands or is compressed is heated or cooled according to the same law.

With SPH comes the ability to choose the manner in which one
incorporates the isothermal evolution. We may set the temperature 
of each particle as a function of its distance to the star (Eulerian
implementation), or we may keep its temperature fixed no matter
where the particle moves (Lagrangian implementation). In most of our
simulations, we have chosen to use the Eulerian version.  This choice
is dictated by consistency, since the isothermal assumption implies
that the star must contribute the bulk of the heating, and by the
desire to match as closely as possible the PPM calculations.

\section{Results of Simulations}\label{results}

With the initial conditions outlined above, we have run a series of
simulations with both codes in which we vary disk mass but keep a
constant minimum Toomre parameter \qmin~$=1.5$. A free outer boundary
condition was implemented for each SPH simulation, while one series
of PPM simulations was run with a reflecting outer boundary. A second
series of PPM simulations used an infall through the outer few cells
onto the outer disk edge, which was assumed to be an initially 
stable isothermal shock.  To explore varying stability, we also ran
two SPH and one PPM series varying \qmin~between $1.1$ and a
maximum value defined by the condition that the outer edge of the 
disk remained bound. Each simulation was evolved for periods ranging
between a small fraction of an orbital period $T_D$ (in the case of
very low \qmin~runs in which rapid clumping was seen), to several complete
orbital periods for runs in which clumping was not observed. 

Unless otherwise specified, no explicit initial perturbations have been 
assumed beyond computational roundoff error. Due to the discrete 
representation of the fluid variables, this perturbation translates to
a noise level of order $10^{-3}$ for the SPH calculations. The relatively 
large amplitude of the noise originates from the fact that the hydrodynamic 
quantities are smoothed using a fixed number of neighbors (see Herant \&
Woosley 1994). An increase in the number of particles does not necessarily
decrease the noise unless the smoothing extends over a larger number of
neighbors. Because of its similarity to Monte Carlo methods, the decrease
in noise goes as the square root of the number of neighbors, and so
decreases slowly with a large increase in computational cost.

For PPM, the noise level can be made as small as machine precision 
(while  double precision is used internal to the code, single precision is 
used in initialization and dumps, so we obtain $\sim 10^{-7}$).  The PPM
simulations are terminated at a perturbation amplitude of 
$\delta\Sigma/\Sigma \sim20$\% because matter on elliptical orbits begins to
interact strongly with the inner and outer boundaries. SPH simulations on 
the other hand, are carried out until clumps begin to form
(clumping causes the time step to drop drastically and halt the evolution).
Highly stable disks, for which clupms do not form, are terminated when
no significant additional evolution is anticipated. Each of the SPH 
simulations run for much of their duration with high amplitude
($\delta\Sigma/\Sigma \sim100$\%) perturbations. Comparison simulations 
on a simple test problem (see section \ref{sphppm} below) were run to 
high perturbation amplitude using both SPH and PPM in order to confirm
the late time behavior of the SPH simulations.

We did not formally introduce a perturbation in our
initial conditions, however two conditions provide indirect seeds
for perturbations. First, the disk is cut off at an inner
radius which, while small, is nonetheless large compared to the
stellar radius. This cut off creates a gravitational potential hump at the
center, and is equivalent to a strong seed for the $m=1$ disturbances.
As the star moves away from the origin, it is further
accelerated by the hump, effectively sliding down
the incline.  We show in figure \ref{gravpot} the gravitational potential 
near the origin for the disks with the characteristics described above 
as well as the tori used in our comparison calculations below (section 
\ref{sphppm}).  By following the procedure of Heemskirk \etal (1992),
who derive an equation of motion for the star including the zeroth
order hump term plus first order perturbations, we note that initially
the growth rate for a $m=1$ pattern will be
\begin{equation}
\gamma_1 = \left(\sqrt{ {{d^2\Psi_D}\over{dr^2}} }\right)_{r=0}.
\end{equation}
Computing numerically the curvature of the hump, we derive a $m=1$ 
growth rate due to the hump of $\gamma_1/\Omega_D\sim 5$. Indeed during
the very earliest stages of our simulations ($t\lesssim 0.1T_D$), we
find a growth rate of this (quite large) magnitude. After the
initial transient, growth rates quickly fall to more sedate levels.
The contribution to the long term global growth of instability due to
this initial perturbation is thought to be a small component of the
total. 

A second indirect seed of dynamical instability is connected to the
fact that the density law has been softened (eq. [\ref{denslaw}]) or
modified (eq. [\ref{denslaw-a}]) in the innermost regions of the disk
in order to avoid a singularity at small radii. This density decrease
creates a region of high vortensity gradient which excites excites wave
growth (see Papaloizou and Lin 1989). This instability channel also 
requires a seed, but its proximity to the inner edge, where orbit 
times are small, coupled with the hump perturbations, make the time 
scale for its initial excitation quite short.

Features of our simulations are tabulated in Tables \ref{sph-tabl} and 
\ref{ppm-tabl}.  The first column of each table represents the name of 
the simulation for identification. The second column defines the 
resolution (in number of particles or grid size). Initial disk/star
mass ratio and minimum $Q$ are given in columns 3 and 4, while total
simulation time and spiral features of each simulation fill out the
remaining columns.

We illustrate the phenomena seen in our simulations by presenting two
representative cases: mass ratios \mrat~$=1.0$ and \mrat~$=0.2$. Both use
initial values of \qmin~$=1.5$. These disks represent points near either
end of a spectrum of behavior. In section \ref{tempvar}, we show additional
models which vary \qmin, demonstrating behavior along another axis in 
parameter space. We first examine the qualitative nature of the simulations,
then examine in detail the structures which form.  A comparison of
the results of SPH and PPM and limitations imposed by numerical features
is discussed in section \ref{sphppm}.

\subsection{General Observations and Morphology}\label{genobs}

Spiral arm growth occurs with varying rates and amplitudes. Growth is not 
smooth or continuous. Frequently arms change shape, stretch, or break off 
and drift until hit by another passing disturbance. Well developed spiral
arms, while subject to irregular change on short time scales, do survive.  
In figure \ref{sph-himas}, we show a series of snapshots of particle positions
in simulation {\it scv6}, characterized by \mrat~$=1.0$ and an initial minimum
\qmin~$=1.5$. Instability first develops in the central regions of the disk,
and propagates outward in radius. Even early ($0.5 T_D$) in the simulation,
variations of density $\delta\Sigma/\bar\Sigma$ approach 10-50\%; at late 
times they reach unity.

The dominant patterns are two and three-armed spirals with significant
components having other symmetries. At late times we see multiple tails on 
a single arm, arms unevenly spaced in azimuth, and patterns which 
have one arm which is significantly stronger than its counterparts.  
Often such spacing and asymmetry is preceded by the breakup of an arm 
at its base, and subsequent drift through the disk or capture by another 
arm. For example, between the $0.94 T_D$ and the 
$1.41 T_D$ images, an $m=2$ structure breaks up, and reforms as an
asymmetric $m=3$ spiral pattern. It then returns to its previous two
armed structure by $1.73T_D$.

A comparable series of plots for a PPM simulation ({\it pch6}) with
analogous initial conditions is shown in figure \ref{ppm-himas}. The 
variable plotted is density variation defined by
\begin{equation}\label{densvar}
{\Sigma_{ij}-\bar\Sigma_i}\over{\bar\Sigma_i}
\end{equation}
where $i$ and $j$ refer to the grid indices of the $r$ and $\phi$
coordinates respectively, and $\bar\Sigma_i$ is the azimuth average 
of the surface density at radial grid index $i$.  Only positive 
variation contours are shown. The linear spacing between
one contour and the next higher contour is noted in the upper right
hand corner of each plot. The dotted line denotes the disk edge at
50 AU. As in the SPH simulation, instability begins in the inner 
regions of the disk. Complex structures follow at midtime epochs. 
Later behavior shows well defined regular spiral patterns,
with a mix of several patterns dominated by $m=2$ and $m=3$ which 
dynamically reorganize themselves with time.

The simulations above display a number of similar characteristics, though 
on a different spatial scale and mass distribution, to the protostellar 
core/inner disk simulations presented in Pickett \etal (1998). In each case, 
large scale spiral structures grow from marginally stable systems. The 
instabilities begin their growth in the innermost regions of the system and 
proceed to involve the entire disk as the simulation proceeds. At late times
in both sets of simulations, the spiral arms become azimuthally condensed. 
A notable difference between our results and theirs, which will be discussed 
in section \ref{disk-grw} below, is the fact that our simulations exhibit a 
pattern speed which increases toward the center of the disk. In contrast, 
Pickett \etal report constant pattern speeds.

Initial behavior of our low disk mass runs is similar to those of high mass,
with instabilities first becoming apparent in the inner regions of the disk. 
Evolution at later times differs from that for high mass disks. We see the 
rapid development of patterns with large numbers of spiral arms, which 
display a tenuous, filamentary structure not present in higher mass disks. 
The disk shown in figure \ref{sph-lomas} (simulation {\it scv2}) has a five 
armed pattern which predominates, and at late times fragments into multiple 
clumps from each arm. A region of apparent stability against spiral arm
formation  becomes apparent in the extreme innermost regions (see also
section \ref{tempvar}).  Such regions are present to some extent in 
all of our SPH disks except those which form clumps immediately and are
defined by a value of Toomre's $Q$ parameter greater than $\sim$2.

In a low disk mass PPM simulation ({\it pch2}), shown in figure 
\ref{ppm-lomas}, we also find a change in character and an increased
number of spiral arms. As in figure \ref{ppm-himas}, the instability 
begins to form its first spiral structures at amplitudes of 0.01-0.1\%.  
Although the precise number of arms seen does not correspond to that 
in the SPH run (showing instead the 2-4 armed patterns dominant), the
degree of small scale fragmentation in the region around 5-25 AU is
similar. We believe that the partial suppression of the high $m$-number 
patterns can be attributed to the wavelengths of those patterns 
approaching the gravitational softening length implied by the grid. This
statement is supported by the fact that for the low mass disks ({\it pch2}
and {\it pcm2}), the amplitude of the perturbations $\delta\Sigma/\bar\Sigma$, 
is larger in the higher resolution simulation. These simulations do not
resolve the small scale structure. Note that the PPM run with \mrat~$=0.1$
($pcm1$) developed only minimal spiral structure after nearly six full 
disk orbit periods.

Structures observed in the moderate resolution PPM simulations (runs 
{\it pcm1-pcm6}) were qualitatively similar to those observed for
our highest resolution runs ({\it pch2, pch6}), although the growth of
the low mass/low resolution structures was slower. Growth rates are 
similar for the low and moderate resolution high mass disks.  
The simulations may have reached a level of convergence sufficient 
to resolve the large scale features of the evolution, but further 
improvement is desirable.

\subsection{The Effects of Temperature} \label{tempvar}

Two series of SPH simulations were run varying the minimum stability
parameter $Q$, with mass ratios \mrat~$=0.8$ and $0.4$. Other things
being equal, high $Q$ implies high temperature in the disk (eq.~[\ref{Qdef}]). 
We vary $Q$ for different simulations between a minimum value of $Q=1.1$, 
at which the disk is only marginally stable to ring formation, and a
maximum value such that the outer edge of the disk remains bound. 
For the high mass series, this limit was found at \qmin~$=2.3$, while for
the lower mass disks up to \qmin~$=3.0$ were available.

In figure \ref{sph-qvar} we show `late time' behavior of each of the disks 
in the \mrat~$=0.8$ series ({\it sqh1-sqh6}). Below an initial value of
\qmin~$\sim1.4$, strong instability and clump formation occurs during a few 
orbit periods of the inner portion of the disk. The outer disk remains 
largely unaffected during the simulation (which suffers drastic decreases 
in time step size once clumps form). At moderate \qmin~(1.4 to somewhat 
less than 1.7), instability in the inner regions is slowed to the extent 
that spiral instabilities involving the entire disk have time to grow. 
These spiral arms then become filamentary and clump on time scales of one
or two $T_D$.  The last few frames in figures \ref{sph-himas} and 
\ref{sph-lomas} show such behavior for a disk with \qmin~$=1.5$
and \mrat~$=1.0$ and $0.2$. The portions of the spiral arms at large
distances from the central star remain thicker and more diffuse,
while the inner regions evolve toward more sharply defined features.
As \qmin~increases the character of the spiral instabilities changes
from narrow, filamentary structures and clumps in the inner disk 
to thicker arms which develop on disk orbital time scales at higher
initial \qmin.

Above initial \qmin~$\sim2.0$, we see only limited asymmetry and spiral
structure. However, there is a strong transient epoch in which the 
centers of mass of the star and the disk orbit each other at large 
distances (relative to their late time behavior or to other, less stable
(lower \qmin) simulations). Simulations have been carried out to more 
than $4 T_D$ for these cases. This resonance gains in strength with 
increasing \qmin~up to the maximum values simulated. Accretion of disk
matter onto the star occurs at higher rates in these runs as well.
The star makes a hole in which little disk matter remains. 

Figure \ref{hiq-trans} shows an example of this transient for simulation 
{\it sqh6}.  The orbit of the star begins with a slow transient to
relatively large distances from the system center of mass (as large as
$\sim0.05 R_D$ in the disk shown). In the first $\sim 2 T_D$,
the star accretes a large fraction initially located in the inner part
of the disk. After this time the star settles to a smaller orbit, with 
occasional fluctuations  as it moves in response to disk perturbations, 
and continues to accrete from the inner disk. We believe this  
transient is largely due to the high mass accretion rates with
nonaxisymmetric flow.  With such fast accretion, the flow of mass
onto the star is rapid enough that appreciable angular momentum is
swept along as well. A comparable simulation, with the star fixed at
the origin, shows nearly as large an accretion rate. We conclude that
high accretion drives the stellar migration, rather than the reverse
process where the star moves by some other means (caused for example
by a torque from the outer disk) into a region of the disk in which 
high accretion may take place.

Although we find that the accretion rates seen in the most $Q$-stable
disks are higher than low stability disks, it is not clear whether the
magnitude of the accretion rates are correct. In SPH the accretion 
of a particle implies a sudden unphysical loss of pressure support to 
the neighbors of the accreted particle. As the disk reorganizes itself, 
additional particles move inward towards the accretion radius. If the 
mass accretion for all of our disks were to be scaled up or down by a 
common factor, the transient in figure \ref{hiq-trans} might increase or 
decrease in magnitude or even disappear. What we can say with certainty is 
that if a star can accrete matter from the inner disk quickly enough that 
it loses its pressure support further out, accretion of disk material which 
has not lost all of its orbital angular momentum can occur, driving the 
star away from the system center of mass. In the simulations we study
here, such a condition occurs when the accretion rate is above
$\sim 6-8\times 10^{-5} M_\odot$/yr for the high mass series and 
$\sim 2\times 10^{-5} M_\odot$/yr for the lower mass series.

Behavior of the lower disk mass series of SPH simulations with varied 
\qmin~is similar. The overall characteristics of the evolution mimic that
of the higher mass runs but are `stretched' along the $Q$ axis to higher
values of \qmin. Azimuthal condensation of spiral arms is again seen up 
to initial \qmin~$=1.5$, but the \qmin~$=1.7$ run at this mass ratio 
appears to be just beyond the critical stability for clumping: many
preliminary characteristics of clumping such as well resolved spiral
arms and short duration over-density spikes (see below) were evident
but no actual formation occurred at the time we stopped the run at
$T=5 T_D$. Production of thick arms continues as high as 
\qmin~$=2.3$ and global star/disk resonances again manifest themselves
all the way up to the maximum \qmin~values studied.  Distinct
one-armed spiral waves form at \qmin~$=2.0$ for short periods, then lose
coherency and fade back into a smooth, global pattern.

One series of PPM simulations was run with varying \qmin. The
late time density variation contours for the series are shown in figure 
\ref{ppm-qvar}.  Because of the low amplitude of the initial noise, these
simulations were continued to $\sim 2 T_D$ even for the lowest 
minimum $Q$ values. In the highest stability (\qmin~$=2$) simulation, 
we find that the strength of the instabilities near the inner boundary 
dominate the instabilities over the disk as a whole. This instability 
does not seem to be the same as the transient seen in the SPH
runs: it is limited to small radii inside the 
density maximum, and does not enter the outer disk at all. Because
of the boundary behavior noted above, simulation of the disks into epochs
having large amplitude variations was possible for only short times.
We could not determine if a large transient in the orbits of the
centers of mass of the star and the disk developed at late times for 
these simulations.

At low and moderate \qmin~($\leq1.7$), there is a great deal of 
correspondence between the qualitative results of our SPH and PPM runs. 
For simulations with moderate initial \qmin~($\sim1.4-1.7$) multiple
spiral arm structures develop with the $m=2$ and $m=3$ patterns 
most prominent. The $m=1$ pattern is present at varying levels
as an asymmetric component of the dominant $m=2$ or 3 patterns.

For the lowest stability simulation, run at \qmin~$=1.1$, density 
variations up to $\sim$40\% are present in the disk and variations
within a single spiral arm produce local density maxima within
that arm. Continued collapse from large amplitude spiral structure into
one or more clumps is not observed, probably because 
we have not resolved the gravitational potential 
or the rotational motion of the matter about a collapsing core to 
the necessary scale.  The evolution of these lowest stability disks
(i.e. simulation {\it pqm1} and {\it pqm2}) at early times in the
simulations are dominated by the growth of the $m=1$ pattern which,
unlike their more stable cousins, is distinct even at the 
$10^{-5}-10^{-6}$ level. Later, these patterns tend to break up and 
reform as $m=2$ and $m=3$ patterns.

\subsection{Spiral Pattern Growth}\label{patterns}

An important connection of numerical simulations to linear perturbation
analyses is to define, if possible, the linearly growing spiral patterns
of a system. To do so requires a specification of the growth rates
and pattern speeds of the dominant spiral patterns in each system.

We compute the growth rates by first computing the amplitude of spiral
patterns by Fourier transforming a set of annuli spanning the disk in
the azimuthal coordinate. The amplitude of each Fourier component is
then defined as $|A_m|=\ln (|\Sigma_m|/|\Sigma_0|)$, where $\Sigma_m$ is
\begin{equation} \label{mode-amp}
\Sigma_m = {{1}\over{\pi}}\int_{R_i}^{R_o}\int_0^{2\pi}
		e^{im\phi}\Sigma(r,\phi)rd\phi dr,
\end{equation}
for $m>0$ and the inner and outer radii of the disk are defined by $R_i$ 
and $R_o$. The $m=0$ term is defined with a normalization of $1/{2\pi}$. 
With this normalization, the $\Sigma_0$ term is the mass of the disk and 
the amplitudes, $A_m$, are dimensionless quantities. The phase angle 
is then defined from the real and imaginary components of the amplitude
\begin{equation}
\phi_m = \tan^{-1}\left[{{Im({\Sigma_m})}\over{Re(\Sigma_m)}}\right].
\end{equation}
Local amplitudes for each component can also be derived for annuli by
neglecting the integration over radius in eq. [\ref{mode-amp}]. Each
Fourier component is computed about the center of mass of the system.

Assuming strictly linear growth for each Fourier component, we can use least 
squares techniques to fit a growth rate, $\gamma_m$, to each amplitude as
a function of time with the equation
\begin{equation}
A _m = \gamma_m t + C_m,  \label{grwtheq}
\end{equation}
where $C_m$ is an constant defining the initial amplitude of the component.
If we keep track of the number of times, $N$, a pattern has wound past 
a phase angle of 2$\pi$ and add $2\pi N$ to the derived phase at each
time, we can derive a pattern speed by a similar fit as
\begin{equation}\label{phidot}
\phi_m = {{\dot\phi_m}\over{m}} t + \phi_{m,0}.
\end{equation}
This definition effectively averages over all short term variations (if
any) in the pattern speed. A periodogram analysis gives similar results to
this fit technique.  The frequency with which we produce dumps of the 
simulation is sufficient to produce accurate pattern speeds over all but
the inner $\sim 3-5$ AU of our disks, and over the full radial extent of
the tori (see section \ref{sphppm}).

We may independently derive an additional global growth rate for the
$m=1$ component by noting that it is the only component which can
contribute to the motion of the star. All higher order components are
symmetric under a rotation smaller than 2$\pi$ radians (i.e. $2\pi/m$
respectively for each Fourier component) and therefore do not contribute 
to the motion of the disk center of mass. By fitting the distance between
the centers of mass of the star and disk as a function of time, we find
a growth rate independent of the precise geometry of the spiral arms in 
the disk. In general we find good agreement between this growth rate
and the value derived from the above procedure.

The analysis of the pattern growth in disks and tori can proceed
at either a local or a global level by either including or excluding
the integration over radius in eq. [\ref{mode-amp}]. If we derive a 
growth rate and pattern speed in a succession of narrow rings in the 
system and compare the values over the entire system, we can readily 
identify structures which are coherently growing and moving over large 
temporal and spatial scales. This feature is limited in a global analysis
because the integration effectively averages the amplitude and speed of
a given pattern over the entire system.

On the other hand, a local analysis can be quite misleading. If we consider
a series of concentric narrow rings making up a disk, we must account 
not only for the growth of instability within any given ring, but also
for the transport of already formed instabilities from one ring to 
another. For example: if some `lump' of matter grows in one ring in the
disk, then moves by some process to a second, the amplitude of the 
Fourier components in each of those rings will be affected: one will exhibit 
a net loss in amplitude, while the other a net gain. A growth rate based 
upon amplitudes affected by such processes would no longer represent
the physical instability mechanisms present in the disk. 

In the analysis that follows, we shall use a local analysis to
identify patterns which are growing coherently over large spatial
scales, but in order to compare our results to the global analyses
of ARS and STAR, we shall utilize globally integrated quantities.

\subsubsection{SPH and PPM: A Direct Comparison of Results 
and Numerics}\label{sphppm}

Each code does well with different aspects of the evolution of 
disks.  For the example of the disks discussed here, the low noise in
the PPM calculations allows an accurate growth rate calculation,
but with our treatment of boundaries, problems develop as a simulation 
becomes nonlinear. Matter reflected from the boundaries changes the total
momentum of the system to such an extent that its center of mass 
(exhibited particularly in the position of the star) attempts to move to
infinity. Because of its ability to dynamically adapt the available 
resolution to the interesting parts of the flow and relative sensitivity
to boundaries, SPH is able to follow the nonlinear evolution much further.
These same features however, forbid simulating a disk with a low density
central hole because the steep density gradient near the inner disk edge
cannot be adequately resolved at a computationally affordable level.  
Even for disks without a hole (for which the gravitational softening at 
the inner boundary blurs the physics and allows the simulation to proceed),
the initial noise in SPH (of order $10^{-3}$) leaves very little time
for random perturbations to organize themselves into ordered global
spiral structures while remaining in the linear regime. 

Fitting growth rates to the SPH simulations requires much more
caution than is required for the PPM runs. The initial noise level is
such that only a very short time baseline is available prior to
saturation.  Typically, we observe a period during which Fourier 
components grow linearly until reaching a saturation level. This
period of linear growth lasts for about one disk orbit $T_D$ or
less for SPH and 2--3 $T_D$ for the PPM simulations.

The SPH disk simulations often reach high perturbation amplitudes close
to the star before more distant regions of the disk have become active.
To compare the two numerical methods and minimize this time scale problem,
we have simulated relatively narrow tori. Such tori have a much more 
restricted dynamic range than a disk, so that the entire system becomes 
active at once.  We use a torus with an outer to inner radius ratio of 
$R_i/R_o=5$ and a $\gamma=1$ equation of state given by eq. [\ref{eos}] 
with temperature, density and individual particle mass given by a Gaussian 
function of radius \begin{equation}\label{torus-law}
f(r) = f_0e^{ -\left({{r-r_0}\over{R_w}}\right)^2},
\end{equation}
where $r_0$ is defined at the midpoint, $r_0=(R_i+R_o)/2$, of the
torus and $R_w=(r_0-R_i)/2$, so that the torus extends about three
`standard deviations' in either direction from the highest density
point (figure \ref{torus-init}). Each simulation is then evolved
isothermally in the same way as is done with our simulations of disks.

With a $\gamma=1$ equation of state, it is difficult to find toroidal
configurations which are initially stable to axisymmetric perturbations
(i.e. $Q>1$ everywhere), except for relatively low mass tori. 
For a variety of temperature or density laws, either the high density
central region will collapse (i.e. the initial \qmin~will be less than 
unity), or the outer edge will be unbound. For our test problem, a ratio
of $M_T/M_*=0.2$  yields a minimum $Q$ of about $1.05$ near the center 
of the torus.  As before, the star mass is $M_*=0.5M_\odot$, the outer 
torus radius of $R_o$=50~AU, and thus the outer edge of the torus orbit 
period is $T_T=T_D=$500 yr.

Table \ref{cmp-tabl} summarizes the characteristics of the simulations. 
The linear and nonlinear regimes are divided by the condition that the 
amplitudes of Fourier components other than the dominant pattern (or 
patterns) reach comparable amplitudes to that dominant pattern, and total
perturbations reach $\sim$10\%.

One SPH and two PPM simulations were run with this toroidal
configuration at a resolution of $40\times150$ cells for the PPM
runs and $6998$ particles in the SPH run. One PPM simulation with
initial random noise amplitude $10^{-3}$ (comparable to the initial
noise in SPH) and one with noise of amplitude $10^{-8}$ were run. 
The $10^{-3}$ noise is input as a random density perturbation in each 
cell as
\begin{equation}
\Sigma_{ij} = \left(1 + 10^{-3}(2R-1)\right)\Sigma_{ij}
\end{equation}
where $i$ and $j$ refer to the radial and azimuthal grid indices
and $R$ is a pseudo-random number between zero and one.
The $10^{-8}$ amplitude noise is derived from truncation 
error in the initial state. Boundary conditions are identical to
those used in our disk simulations.

The relatively large amplitude of the noise in the SPH simulations 
is caused by smoothing over a finite number of neighbors (see Herant
and Woosley 1994).  Increasing the number of neighbors used
in the interpolation has a small effect in decreasing the noise amplitude
but at a high computational cost. We have used a varying number of
neighbors (depending on local conditions of the run) with a distribution
centered near 15--20 neighbors per particle, a number which is standard 
for two dimensional simulations. 

The resolution of features within the torus or disk must inevitably
be less accurate in a finitely resolved system than in a physical
system.  PPM spreads shocks over at least two cells, for example, while
further loss of resolution may come from the representation of the 
gravitational potential.  In SPH, resolution is limited by
the smoothing length of the particles and the artificial viscosity 
required to adequately reproduce shocks. 

Two additional PPM and two additional SPH simulations of tori have been
run to test resolution. One PPM run has 1.5 times the resolution in each
dimension (60$\times$225--roughly doubling the number of cells) and the 
second twice the resolution (80$\times$300--quadrupling the number of cells).
The SPH simulations increase by a factor of two and a factor of four the 
number of particles in each simulation. Comparing runs of different
resolution is difficult, however, because the power spectrum of the initial
perturbations  may not be controlled to the limit required. In an attempt
to duplicate the perturbation at low and high resolution, but remain above 
the uncontrollable level imposed by the grid itself, we have input an initial
random noise amplitude of $10^{-3}$ in each 2$\times$2 block of cells in
each of the two higher resolution PPM runs.

We show the evolved configuration of each run in figure \ref{tor-cmp}.
The time at which each is shown is near the linear regime cut off
discussed below. The SPH runs are mapped onto a grid and plotted in the
same manner as the PPM runs in order to make the visual comparison as
direct as possible. In each of the runs, instability growth is dominated
by $m=2-4$ spiral patterns with the higher resolution runs tending to show
progressively less of the $m=4$ pattern and more of the $m=2$ pattern. 
The $m=3$ pattern predominates in each simulation except for the two low 
resolution PPM runs. The change in morphology in different simulations is 
probably an artifact of the resolution. As we show for the growth rates 
below, the lowest resolution simulations are apparently not converged.

In comparison, the results of Laughlin \& R\'o\.zyczka (1996) show a 
dominance of an $m=2$ component without a large presence of other 
patterns. The origin of instabilities in their systems is attributed to 
the family of vortensity instabilities with corotation exterior to the 
torus. Different initial conditions seem to be responsible for the $m=2$
rather than $m=3$ dominance. Our test simulations use a narrower torus
than theirs, with an isothermal rather than adiabatic equation of state. 
A simulation with an identical initial condition and equation of state  
compares favorably to their results.

The amplitudes and fits for growth rates for the $m=3$ spiral 
pattern at the center of the torus (at $R=30$~AU) are shown 
for each simulation in figure \ref{torm3_30}. The fit parameters are 
derived from only the portion of the curve in which the patterns are 
growing and little disruption of the large scale structure of the tori 
has begun. This disruption is characterized by an onset of fragmentation
at the inner and outer edges of the torus (SPH) or significant radial 
distortions in the torus (PPM). We also allow a short period 
($\sim 0.1T_D$) prior to the first fitted time point, for some
initial transients (eg. the unphysical `ringed' structure in the
SPH initial state) to settle.

The pattern speeds and growth rates for the $m=1-4$ patterns
are shown in figure \ref{ppmpatgrw1-4} for each of the PPM simulations
and in figure \ref{sphpatgrw1-4} for each of the SPH simulations. The
pattern speeds for the $m\geq 2$ patterns for each of the runs agree
for both codes over the range of resolution and initial perturbation
amplitude. The growth rates from the SPH simulations differ by as much
as 50\% between runs. For the SPH simulations obtaining a constant rate 
across each ring in the torus was not possible.  For the PPM simulations,
the growth rates near the inner and outer boundaries of the tori are reduced 
due to the fact that perturbations there do not begin to grow until after 
the denser regions of the torus have been disturbed. A similar effect is 
found for the pattern speed near the inner edge.

The growth rates for the SPH runs are affected by the high amplitude of
perturbations in the initial state and the short time span over which
the fit must be derived. Longer lived initial transients caused by the 
excitation of multiple eigenmodes of the system or by small inhomogeneities 
in the initial state can cause the amplitude curves to become quite 
nonlinear in form. The PPM simulations have longer time baselines so such 
transient effects are less important. 

The growth rates for the $m=4$ pattern in the PPM simulations decrease
with increasing resolution, while the $m=2$ and 3 growth rates are less
affected. This fact and the trend towards $m=2$ and 3 spiral patterns
for higher resolution runs suggest that they may be true linearly growing 
patterns for the system. The change in character with increasing 
resolution may be due to the fact that the torus begins its life very close 
to the stability limit, \qmin~$=1.0$. Any inaccuracies in the resolution 
of the gravitational potential or the mass distribution (hence the pressure) 
will have their greatest effect in such a circumstance. The SPH simulations 
show no comparable effect, but reliable local growth rates can not be 
obtained for those simulations.

Late in each simulation the tori collapse into several condensed objects, 
but the details of the collapse vary. Not all of the spiral arms present 
during the growth of structure condense into separate objects. In many cases 
the spiral arms break up and/or merge as clumps begin to form. Figure 
\ref{tor-late} shows snapshots of each of the runs at the time at which the
spiral arms begin to collapse. Each simulation is halted at this point because 
of the influence of the boundaries on the simulation and because we did not 
properly simulate the physics important in the collapsed objects. The structures 
which develop resemble the simulations discussed by Christodoulou \& Narayan
(1993) because the tori tend to deform radially as instabilities grow. With
both codes the torus becomes so distorted radially that a line of condensations 
forms from the torus matter which has moved outwards.

We now summarize the similarities and differences between the 
results of each code.

Each code produces instabilities which grow in the tori as 
they evolve forward in time. The instabilities produced are 
multi-armed spirals structures which, at the end of each simulation,
have begun to radially distort the torus and collapse into clumps.
In both codes predominantly 2--4 armed spiral structures are produced.
The high resolution simulations each produce 2 and 3 armed structures 
while low resolution simulations (apparently incompletely converged), 
produce predominantly 3 and 4 armed structures.

The initial state of an SPH simulation begins with random noise of
amplitude $\sim 10^{-3}$ above or below an `ideal' initial value.
Near the boundaries, where particles are not distributed evenly
with respect to each other, additional differences from an ideal
initial state are present. PPM can begin with noise in the initial
state as small as machine precision for any given simulation.

The differences between one code and the other can be attributed to
several effects. First, perturbations in the initial state may
trigger more than one true eigenmode of the system which, taken
together, cause more or less observed growth in a given simulation 
with respect to another. Because the noise input for each code 
arises from such different sources, the stimulated pattern growth
may therefore initially have a much different  character.  This
growth rate variation is exhibited predominantly by the amplitude
of a given pattern `waving' above and below its true linear growth
curve and, in essence, constitutes an error estimate for a calculated
growth rate. The PPM simulations, for which the growth rates are 
calculated over longer time baselines and with a smaller initial
noise amplitude per Fourier component are not nearly as strongly
affected by such effects. We estimate errors of 10-20\% in the
growth rates due to this effect in the PPM simulations and perhaps
an additional 20-40\% in the SPH runs because of their very short
time baseline. Pattern speeds do not seem to be as strongly 
affected by these transient effects.

The adaptive nature of the resolution and high noise in SPH causes
small scale filamentary structures to become active and develop more
quickly than in our counterpart PPM simulations, which are limited
to the resolution of the fixed grid. SPH will tend toward developing
grainy and filamentary structures quickly, perhaps to a larger extent
than is physically the case.

Because the grid boundaries are far away from the main concentration
of mass in the torus, they have only a small effect until late in
any given simulation. Such is not true for the disk simulations
using the PPM code so those simulations cannot be carried out far 
into the nonlinear regime due to the growing influence of the 
boundaries at late times. The physics important for the global 
dynamical evolution of the disk ranges over a dynamic range larger 
than we are able to simulate. The state at which the PPM runs must 
be terminated (with 10-20\% perturbations) are qualitatively quite 
similar to those of the SPH runs over most of their duration. 
It may be that for the disks we discuss below, the PPM runs are 
representative of the linear regime, while the SPH simulations are 
our only representation of the late time nonlinear behavior of the 
system.

\subsubsection{Pattern Growth in Disks}\label{disk-grw}

With a clearer understanding of the numerical properties of our 
codes on a test problem, we return now to the study of disks. 
Due to the high initial noise of the SPH runs and large radial 
extent of the disks we study, saturation at small radii often occurs
well before the entire disk has become involved in the instability.
Because of this noise we do not believe growth rates calculated from these 
simulations are reliable for any Fourier component except the globally
integrated $m=1$ pattern (for which we have the behavior of the centers
of mass of star and disk), and we limit discussion of the growth rates 
in this and the following sections to the PPM simulations.

The qualitative observations of sections \ref{genobs} and \ref{tempvar}
have shown that there is rarely a single spiral pattern present in a disk.
More quantitative measurements show that growth is present in all 
Fourier components up to very high order. Such growth does not necessarily 
imply that actual spiral arms of that order are present in the simulation, 
but rather that the arms that do exist become more filamentary than pure 
sinusoids, creating power in higher order Fourier components (a Dirac
$\delta$-function will yield power at all wave numbers for example).
In order to be more definitive regarding the true morphology of each 
disk we visually examine each simulation and tabulate the dominant 
spiral patterns in Tables \ref{sph-tabl} and \ref{ppm-tabl}.

Which patterns represent linear growth in each of the systems? 
To begin to answer this question we must fit growth rates and pattern
speeds to the various spiral patterns present in each disk and
determine which patterns exhibit rates which are constant at differing
resolution, across a large portion of the system and over a large
time period. In figure \ref{m2and3hi} we show the amplitude of the 
$m=2$ and $m=3$ patterns as a function of time near the middle of the
power law portion of the disk and integrated over all radii for our 
prototype massive disk shown in figure \ref{ppm-himas}. Over long periods 
the growth is essentially linear in character. Over shorter periods it 
is punctuated by transients which can change the amplitude by up to an 
order of magnitude. The amplitude variations apparently arise as 
short-lived structures successively grow and fragment throughout the 
disk. Time dependence of pattern speeds within the disk will be discussed
in section \ref{non-lin-phenom} below.

Radius dependent growth rates and pattern speeds for the $m=1-4$ 
are shown in figure \ref{patgrwm1-4hi} for two different grid resolutions. 
The growth rates and pattern speeds are similar at both resolutions, 
suggesting that the simulations may have resolved the physical processes
important in this disk. The growth rates for the $m\geq 2$ patterns are
nearly constant with radius but the pattern speeds derived are not at all 
constant with radius; they decrease as a steep function of the
distance from the central star. 

Low mass disks show a marked absence of the dominant low order ($m=1-3$) 
spiral patterns so common among higher mass disks. Typically, the amplitudes 
and growth rates of all Fourier components are comparable.  We plot the
growth rates and pattern speeds for the same patterns ($m=1-4$) as above
for our prototype low mass disk in figure \ref{patgrwm1-4low}. We again 
find that the pattern speeds are steeply decreasing functions of the radius.
We also find that the growth rates do not exhibit the same values for 
different grid resolutions. This fact suggests that the low mass disks have 
not fully converged at the grid resolution used in our simulations. The 
systematic trend towards faster growth in the higher resolution simulation
indicates that the small scale features which dominate the morphology of
this system may be somewhat inhibited by the resolution of the gravitational
potential and the hydrodynamic quantities on the grid. Much higher resolution
simulations are required to be able to fully resolve the features important
for disks of mass less than $\sim 0.2 M_*$ than are required for more massive 
systems.

With simulations of varying stability we would ordinarily expect 
larger $Q$ values to lead to slower instability growth. Similarly,
we expect smaller $Q$ values should imply more rapid growth of 
instability. In fact, as discussed in section \ref{tempvar}, both extremes
lead to rapid instability growth, but of different character. 

Although it begins with an extreme initial condition, the simulation
{\it pmq5} (with \qmin~$=2.0$, \mrat~$=0.8$) shows an interesting
example of the limiting behavior displayed in a highly stable disk
($Q>>1$ everywhere) with a turnover in its density profile near
the central star.  We show the $m=1$ and $m=2$ pattern amplitudes at two
locations in the disk and integrated globally in figure \ref{hiq-amp}. 
In this simulation, rather than being suppressed, the amplitude of the
instabilities begins to grow quickly in a region limited to the innermost 
portion of the disk. Further out in the disk much slower growth occurs.
The development of such instabilities in disk systems cannot be attributed
to a global, linearly growing phenomenon; its localized character and 
the different behaviors of the amplitude growth at different locations
in the system argue against that. It remains unclear to what extent 
this type of growth happens in real systems, but it seems that with
a turn-over in the density law at small radii or the less severe case 
where the density law flattens (as in our SPH simulations) can lead 
to increased local instabilities. 

It is interesting to note that Pickett \etal (1996) report similar 
behavior (which they refer to as `surge' growth) in several of
their more $Q$-stable simulations. In their work however, the initial 
mass distribution and and rotation curve are somewhat different than in 
our own work. The fact that similar behavior is observed in simulations of
such different character suggests a similar mechanism may be driving the
evolution of both sets of simulations.

The lowest stability simulations also show rapid growth of spiral
instabilities. In these simulations there are no growth features 
similar to the `hump,' or sudden rise in amplitude shown in figure 
\ref{hiq-amp}. In general, the qualitative features of the growth are 
similar to those seen in figures \ref{m2and3hi} and \ref{patgrwm1-4hi}
but with as much as 50--100\% larger growth rates in the case of the
lowest stability run ({\it pmq1}).

The results of our analysis in this section show that in spite 
of its large amplitude at early times and its continued presence
for the duration of the run, our simulations do not show evidence of
a pure $m=1$ pattern. In no case is the $m=1$ growth rate or pattern
speed constant across a large portion of the disk. In contrast to 
several higher $m$ patterns, the wide variation is true of both
the growth rate as well as the pattern speed. Because of the 
variation of the growth rate and pattern speed we must conclude that a 
direct connection to the SLING mechanism is not possible. At the high
amplitude (late time) phase of evolution shown in the SPH simulations, the
$m=2$ and $3$ patterns have become dominant for disks more massive than 
\mrat~$\approx 0.2$, while at the lower amplitudes typical of our PPM runs,
$m=1$ has the largest amplitude, though the pattern itself is ordinarily
seen only as asymmetries in higher $m$ structures.

None of the disk simulations we have performed produce pattern speeds 
for any $m$ pattern which are constant across the entire disk. The 
growth rates, while ordinarily stable at a single value over the whole 
system for at least some patterns (see eg. fig. \ref{patgrwm1-4hi}), do
not reflect the short term behavior of the system as structures 
fragment or deform over time. In this case the `linear growth modes' 
of the system, defined as the complex eigenvalue of a system of 
equations, become difficult to define or to interpret.

\subsubsection{Suggestions for the Mechanisms of Instability 
Growth}\label{mechanisms}

In each of our simulations, instabilities are generated in the 
innermost portions of the disk, eventually impacting the entire
system. Such growth occurs in spite of the fact that the inner regions 
are the most stable as measured by two of the classic stability 
indicators, namely the Toomre $Q$ criterion and the SWING $X$ parameter.
If we are to suggest a mechanism for the instability growth we 
are limited to mechanisms which can produce instabilities in what
are ostensibly highly stable regions.

We have already discussed the possibility that in some cases 
instabilities may due to nonaxisymmetric accretion of disk 
matter onto the star or by accretion of infalling material onto the
disk, rather than to dynamical instabilities in the disk itself. In other
cases, the vortensity based instabilities of Papaloizou \& Lin 1989 
(see also Adams \& Lin 1993) may provide an answer because they can 
grow in highly `stable' regions and their growth can be local in nature.  
They discuss three classes of vortensity instabilities which can exist 
in a disk: those dependent on vortensity extrema within the disk or 
at its edge (`edge modes'), those dependent on resonances (`resonance
modes'), and those which have corotation exterior to the disk 
(dubbed `slow modes' and studied extensively by Laughlin \& R\'o\.zyczka 
1996). Because we find corotation within the disk for most times
(though at varying position), we can eliminate the last of these classes
from consideration. The remaining two, we believe, are both active at 
different times and to a greater or lesser extent in the disks we 
model. At early times, our initial condition (the softened power law
or density turn-over at small radii) implies a vortensity extremum near
the inner boundary of the disk. This condition may excite an edge mode
which over time propagates outward over the density maximum in our PPM
simulations via a resonance mode into the disk, exciting global 
instability channels such as SLING as it propagates into the main 
disk. We have not established a definite connection between the 
instabilities in our simulations and the vortensity based instabilities 
however.

We cannot definitely connect the SLING instability directly to phenomena 
present in our simulations; see  section \ref{disk-grw}. We may still
perhaps be able to make qualitative connections between phenomena predicted
to be important via linear analyses and our results. One example of such 
phenomena would be growth rates which depend upon the outer boundary 
condition imposed. Another might be a growth rate which, as a function
of disk mass, increases for disks more massive than some critical value,
as suggested by the `maximum solar nebular mass' discussed in STAR. Such
characteristics would not necessarily be limited to the $m=1$ pattern
but may also exist in $m>1$ patterns as well. 

We do see such characteristics in the variation of the growth
rates with respect to the disk/star mass ratio. For each series of PPM
simulations varying disk mass, figure \ref{disk-mratrates} shows the 
value of the globally integrated growth rates for the $m=2$ patterns.
Growth rates for other $m$ patterns appear qualitatively similar to those
shown. As one expects, growth rates of the highest mass disks are 
the largest, while instabilities in low mass disks grow much
more slowly. In the reflecting boundary runs, a distinct `turn on' mass 
is evident between $0.2<$~\mrat~$<0.4$, a value which corresponds to the
`maximum mass solar nebula' predicted by the results of STAR.
The infall series does not exhibit such a distinct onset, but rather
a continuous rise to a plateau which does not flatten out until the mass
ratio reaches \mrat~$\approx 0.5$.  

For low disk masses, the growth rates for each pattern are of order 
$\gamma_1/\Omega_D=0.15-0.2$. These rates are comparable to the rate
attributable to numerical effects. The numerical effects have their
origin primarily in the fact that mass interacting with the grid boundaries
gives an impulse to the system center of mass, which must be stable in order
to determine the amplitude of the $m=1$ spiral pattern. Higher $m$ patterns
are also affected as spiral waves reflect off the grid boundaries back into 
the simulation.

For higher mass disks, the outer boundary has a marked influence.  As ARS 
predict, details of the outer radial boundary are an important factor in the
growth pattern. The simulations with matter infalling onto the outer disk 
edge develop spiral structure with growth rates as much as 2-3 times faster 
than with a purely reflecting boundary. Simulations at two resolutions were 
run with an infall boundary to test the degree to which numerical effects of 
the boundary were affecting the growth. Both series show similar growth rates 
(fig. \ref{disk-mratrates}). 

\subsubsection{Importance of Phenomena not Included 
  in Linear Analyses}\label{non-lin-phenom}

On short time scales the pattern speeds in our disks can vary by as much
as 100\%. One example, shown in figure \ref{m2pat_tme}, is taken from the
high mass disk simulation {\it pch6}. There we show the instantaneous 
pattern speed for the $m=2$ pattern near the middle of the disk, as 
calculated by numerically differencing the pattern phase $\phi_m$, at
successive output dumps of the simulation. Such variations in time are
typical of each pattern in each disk simulation we have performed, and
appear in both local and globally integrated pattern speeds. Pattern speeds
calculated this way for the torus simulations of section \ref{sphppm} show 
much slower variations.

In the case of the $m=1$ pattern, whose global pattern speed is reflected 
in the motion of the star, we find that the star occasionally loops back upon
its own trajectory and counter-rotates with the disk for a short period.
Such a condition is not an uncommon occurrence in systems with disturbances
with different orbital (pattern) periods. In our own solar system, for 
example, the sun's motion about the solar system barycenter was retrograde
most recently in 1990, when Jupiter was on the opposite side of the sun from
the other three major planets.

The variations seem to arise because of the growth, fragmentation
and reformation processes undergone by the spiral arm structures
over the course of their evolution. Because the pattern speeds vary,
an averaged pattern speed at any location in the disk (via eq. 
[\ref{phidot}]) loses meaning and the location of the corotation
and Lindblad resonances for each pattern also vary in time. When such
variations are occurring, wave analyses, which typically assume 
stable resonances, may be of limited utility (wave analysis is of
course useful in less chaotic circumstances--see, eg., STAR and
Laughlin, Korchagin, \& Adams 1996).

The growth of instabilities does not diminish as $Q$ increases, but the
instabilities do change character; this change is due to the increasing 
importance of effects not modeled in semi-analytic treatments of disks. For 
the high \qmin~SPH runs, these effects are dominated by the nonaxisymmetric 
accretion of disk matter onto the star. As the star begins to move from the 
center of mass of the system (due to ordinary disk processes or the potential 
hump at the origin), some portion of the accretion becomes nonaxisymmetric. 
In the warmest disks, as much as 10\% or more of the disk is accreted over the 
life of the simulation. Disk matter accreting onto the star sweeps 
along some residual angular momentum which is transferred to the star
either as spin (an effect we neglect here) or as net angular 
momentum of the star about the system center of mass. In these cases,
the star may gain enough momentum to be driven further away from the
center of mass and create power in the $m=1$ pattern.

In the PPM runs with infall, the instability growth can include
a component due to the outer disk edge perturbations. These
may be due to infall itself, or to the fragmentation of the boundary.
Although the linear analyses of ARS and STAR showed that the conditions at 
the outer boundary were important for the evolution of the system, they
were unable to fully model the effects that the boundary can have
on the system (see however Ostriker, Shu, \& Adams 1992).

\subsection{Clump Formation and Characteristics} \label{clump}

Returning now to our SPH simulations, in this section we describe 
several qualitative features of clump formation and evolution in
the disks. Due to the unsteady nature of the spiral instability 
growth and the presence of multiple spiral patterns in the system,
each disk sequentially approaches and moves away from conditions in
which clump formation is likely. These conditions are most readily 
apparent in plots of the minimum $Q$ value in the disk 
and in the maximum over-density in the disk (defined as
$\Sigma(r,\phi,t)/\Sigma(r,t=0)$) with respect to time. 

The value of $Q$ is defined rigorously only for an azimuthally
symmetric disk. Nevertheless, as an indicator of the most unstable 
locations in the disk, we examine its value in nonaxisymmetric systems. 
To calculate its value locally we must first determine the epicyclic 
frequency at each point in the disk. We use the same procedure by which
SPH obtains derivatives of all other hydrodynamic quantities. By 
definition
\begin{equation}
\kappa^2 = {{1}\over{r^3}}{{d}\over{dr}} [\left(r^2\Omega\right)^2],
\end{equation}
so the value $d[(r^2\Omega)^2]/dr$, taken pairwise over each neighbor,
is weighted using the SPH kernel. The result is summed to form a local
value of the epicyclic frequency. 

Plots of maximum over-density and minimum $Q$ are shown in figure 
\ref{odqplot} for our two prototype SPH \qmin~$=1.5$ disks.  Each variable
is a global extremum. As such, the value of one could be determined
from a completely different portion of the disk than the other. However,
after only a relatively small fraction of an orbit time $T_D$, the locations 
of minimum $Q$ and maximum over-density are close, at a position between
about 10 and 30 AU.

After a few orbit periods of the inner disk regions, the over-density 
rises to about twice its initial value (of unity). A slow secular 
trend towards stronger spiral arms over the course of the run follows,
punctuated by one or more sharp, short-duration episodes of very strong 
activity in which density locally increases to 5-10 times.  Over-density 
spikes become more and more frequent as the simulations progress,
finally leading to clump formation. With the one exception \mrat~$=0.4$,
\qmin~$=1.7$ which, as noted in section \ref{tempvar}, appears to lie
on the `boundary' between clumping and non-clumping disks, simulations 
which do not eventually form clumps also do not show these large 
over-density events. We attribute the origin of the over-density 
events in our simulations to the growth of spiral instabilities into a
high amplitude nonlinear regime. In this regime spiral patterns 
present constructively interfere with each other or collide with other 
arms and orphaned arm fragments. 

The results of Adams \& Watkins (1995; hereafter AW) show that a density
enhancement within a disk will lead to collapse if the condition
\begin{equation}
{{\Sigma(r,\phi,t)}\over{\Sigma(r,0)}} > {5 \over 2} Q
\end{equation}
is met, where $Q$ is the local value (azimuth average) of the Toomre 
parameter at the location of the density enhancement. For the disks in
our study, this expression implies that an over-density factor of 3 or
higher must be present in the disk, depending on where in a disk the 
collapse event occurs. This prediction is supported by our numerical 
results, which show that disks can survive (i.e. not exhibit collapse)
for long periods with over-densities of 2-4, but collapse when 
over-density spikes of magnitude 6-10 occur.

For all disk masses, the minimum value of $Q$ rapidly falls below its
initial value to well below unity.  After the initial steep decline, a
slower decrease occurs until clumping begins and minimum $Q$ falls to
zero. The initial decline occurs most quickly in the highest mass
disks, in which instabilities of any type are most strongly felt. 
With $Q$ below unity, the disk becomes unstable not only to spiral
instabilities but also to ring formation or, in the case of isolated
patches, collapse. The collapse is slowed by the effects of rotation
within the forming clump.

We can verify that it is rotation which slows the collapse by noting that
the effects of the over-density spikes manifest themselves at only the 
20-30\% level in $Q$. We also know that the sound speed is constant in 
the proto-clump (due to our assumption that the disk evolves isothermally), 
from the definition of $Q$ we know that the rotation of an individual 
proto-clump (really the shear across the clump, measured by the local 
value of the epicyclic frequency $\kappa$) is the mechanism which inhibits 
further collapse. Only after spiral arm amplitude has reached sufficient 
levels to overcome rotation can an irreversible collapse begin. 

Clumps condense out of the spiral arms on quite short time scales
in even the least massive disks.  During and after the initial
stages of their formation, we find that the clumps show prograde 
rotation. No clumps were seen to form in any disk studied whose
initial \qmin~was greater than 1.5.  Clump formation is most common
at radii less than $\sim0.5R_D$ and usually several clumps will
form from the same disk (and even within the same spiral arm).
Less massive disks form many low mass clumps and higher mass disks
form 2-4 higher mass clumps. The mass inside the clumps is of order
1\% of the star mass at the time each simulation is ended. It is 
clear, however, that from the amount of remaining disk that no 
final mass has been determined.

The clumps form with such vigor in each of these disks because of the
strong cooling implied by the isothermal assumption. Any density
enhancements like those seen in figure \ref{odqplot} instantly lose their
pressure support and collapse rather than dispersing. With more realistic
cooling, the clumping behavior seen in our results may change. Thus our
results are most useful as an indication of the behavior of disk clumping 
and as indicator of where clumps may be most likely to form in more 
physically realistic disks. 

Figure \ref{formrad} shows a plot of the radius at which each clump 
formed for each disk in the series. Only in the case of the 
\mrat~$=0.2$ disk, in which clump formation is prolific in nearly
all regions, were any clumps formed at radii greater than 0.5~$R_D$.
With this exception, we believe the variation in the locations of clump
formation in disks of different mass in figure \ref{formrad} to be due
more to stochastic effects rather than any physical process. To test
this idea we ran a comparison series of simulations ($\times$'s), 
utilizing the Lagrangian version of the equation of state.  When such 
an assumption is made, the background noise inherent in the code 
changes character. No overall structural changes are evident in figure 
\ref{formrad}, but differences in detail are present. Also, for the disk 
with \mrat~$=0.2$, clumps were not formed at the largest radii. We believe
this lack of clumps is due in part to the radial motion of some warmer 
particles into the outer disk, causing clumping to be suppressed.

The prior results of AB92, in which clumps are seen to form at much
larger radii, correspond to a somewhat different initial
configuration. In particular, our present results use a much smaller
`core radius', $r_c$, for the density and temperature power laws. The
gravitational softening parameter for the star is correspondingly
smaller, and no initial perturbations are assumed. These differences
conspire to push collapse instabilities to larger radii in the AB92
results, since in their simulations more mass is concentrated at large
distances from the star. We believe the present conditions to be 
more realistic and thus to represent an improvement over 
the AB92 results.

\subsubsection{Initial Orbital Characteristics}

Out of the entire sample of newly formed clumps, none have an initial
eccentricity much higher than $\epsilon=0.2$, and most are
between zero and 0.1. The low mass companions now being discovered 
around nearby solar type stars show both small and large values
of eccentricity (Mayor \etal 1997; Marcy \& Butler 1995; Butler \& Marcy 
1996). Although the clumps in our simulations form only in relatively low
eccentricity orbits and are therefore dissimilar to many of those being 
discovered, considerable evolution of eccentricity can take place between 
the times corresponding to the end of our simulations and the final morphology 
of the system (see e.g., Artymowicz 1993, 1994; Goldreich \& Tremaine 1980).

\section{Conclusions}\label{summary}

By using two conceptually different hydrodynamic methods (SPH and PPM), 
we are able to simulate a broader range of problems, but gain a sobering 
insight into the limitations of these tools. It is striking that PPM 
indicates violent behavior near the inner boundary (weakly supported by SPH), 
and that SPH indicates pronounced clumping (weakly supported by PPM).
Both methods indicate that instability growth is not a steady progression 
from low to high amplitude perturbations with a single dominant pattern
present throughout. Both methods indicate a marked change in the 
character of instabilities with disk mass. Low mass disks form many
armed filimentary spiral structures while high mass disks form few armed
grand design spiral structures.

In this study of the evolution of circumstellar accretion disks, we
have found simultaneous growth of global spiral instabilities with 
multiple Fourier components. Growth of each of the components occurs 
over the course of a few orbit periods of the disk and a single component
rarely dominates the evolution of a disk. As expected, the massive disks 
are found to be the most unstable, due to self-gravitating instabilities 
within the disk. Accretion of matter onto the star itself can, in warm 
disks (i.e. those with high \qmin~values), significantly drain matter
from the disk on similar time scales to the self-gravitating instabilities.
Short-term variations in the amplitude of a given component, and strong 
constructive interference behavior between different components, can 
produce `spikes' in the surface density. These spikes can eventually 
grow to such amplitude that gravitational collapse occurs resulting
in the production of one or more clumps.

Pattern growth is stimulated at early times by the rapid growth of
instabilities at small radii which eventually engulf the entire disk.
Steady spiral arm structures are not generally present. Instead, spiral
arms progressively grow, fragment and reform as time progresses. In 
cases where accretion is rapid, power can be produced in an $m=1$ spiral
pattern due to nonaxisymmetric accretion of mass and momentum onto the 
star. Understanding the dynamics of the inner region is of primary
importance for understanding the global morphology of the system.

The gross structure of low and high mass disks are markedly different
from each other. High mass disks form large, grand design spiral
arms with few arms, while low mass disks form predominantly thin,
filamentary multi-armed structures. In almost no case is the $m=1$
spiral pattern the fastest growing pattern in the disk. Typically 
a combination of $m=2-4$ patterns in high mass disks or very high
order patterns ($m\gtrsim 5$) in low mass disks dominate the 
morphology. The transition between these behaviors comes at
approximately \mrat~$=0.2-0.4$. This transition corresponds
to the `maximum solar nebula' mass discussed in STAR, above which
$m=1$ modes due to SLING are expected to grow strongly.

It is intriguing to speculate that the collapse processes seen here
are responsible for the production of brown dwarf-like companions such
as that seen by Nakajima \etal (1995) and/or of planetary companions
similar to those recently discovered around several nearby stars
(Mayor \& Queloz 1995, Marcy \& Butler 1996, Gatewood 1996). However, we
must emphasize that clump formation in self-gravitating circumstellar 
disks depends on the ability of the gas to cool efficiently. Our 
simulations here use a simple isothermal equation of state which favors 
clump formation. Additional simulations with realistic cooling functions,
including radiative transfer effects, must be done in order to clarify
this important issue.

\acknowledgments 
We wish to thank the referee, Richard Durisen for a thorough referee
report which improved this paper substantially.  Bruce Fryxell provided
valuable insights into PPM. Greg Laughlin provided valuable discussion 
on the tori we use for our comparisons between SPH and PPM. AFN wishes 
to thanks his collaborators for patience in seeing this work through to 
its completion. This work was supported under the NASA Origins of the
Solar Systems program with grants NAGW-3406 and NAGW-2250. FCA is
supported by an NSF Young Investigator Award, NASA Grant No. NAG
5-2869, and by fund from the Physics Department at the University of
Michigan. DA is supported by NASA NAGW-2798 and NSF ASTRO 94-17346.

\newpage

\include{table-sph}
\newpage
\include{table-ppm}
\newpage
\include{table-cmp}

\newpage
\include{figures}

\end{document}

%% file: table-sph.tex
\singlespace
\begin{deluxetable}{lcccccc}
\tablecolumns{7}
\tablecaption{\label{sph-tabl} Disk Parameters For SPH Simulations}
\tablehead{
\colhead{Name}  & \colhead{Number of} & \colhead{\mrat} &
\colhead{\qmin}  & \colhead{End Time} & 
\colhead{Dominant\tablenotemark{a}} & \colhead{ Number} 
\\
\colhead{}  & \colhead{Particles} & \colhead{} &
\colhead{}  & \colhead{($T_D$=1)} & 
\colhead{Spiral Pattern}  & \colhead{of Clumps}}

\startdata
scv0 &  7997 & .05 & 1.5 & 3.5 & $\gtrsim 12$ & 6  \nl
scv1 &  7997 & .1 & 1.5 & 1.6  & $\sim$10     & 14 \nl
scv2 &  7997 & .2 & 1.5 & 1.6  & 5-6          & 33 \nl
scv3 &  7997 & .4 & 1.5 & 1.7  & 3-4          & 7  \nl
scv4 &  7997 & .6 & 1.5 & 1.7  & 2-4          & 6  \nl
scv5 &  7997 & .8 & 1.5 & 2.4  & 1-3          & 3  \nl
scv6 &  7997 & 1. & 1.5 & 1.8  & 1-3          & 3  \nl
sqh1 &  7997 & .8 & 1.1 & 0.1  & NR           & 18 \nl
sqh2 &  7997 & .8 & 1.3 & .25  & NR           & 11 \nl
sqh3 &  7997 & .8 & 1.4 & .35  & NR           &  7 \nl
sqh4 &  7997 & .8 & 1.7 & 4.2  & 1-2          &  0 \nl
sqh5 &  7997 & .8 & 2.0 & 4.2  & 1            & 0  \nl
sqh6 &  7997 & .8 & 2.3 & 4.2  & 1            & 0  \nl
sql1 &  7997 & .4 & 1.1 & .15  & NR           & 28 \nl
sql2 &  7997 & .4 & 1.3 & 0.3  & NR           &  7 \nl
sql3 &  7997 & .4 & 1.4 & 0.4  & 4            & 1  \nl
sql4 &  7997 & .4 & 1.7 & 5.0  & 1-3          & 0  \nl
sql5 &  7997 & .4 & 2.0 & 4.2  & 1-2          & 0  \nl
sql6 &  7997 & .4 & 2.3 & 4.2  & 1            & 0  \nl
sql7 &  7997 & .4 & 2.7 & 4.2  & 1            & 0  \nl
sql8 &  7997 & .4 & 3.0 & 4.2  & 1            & 0  \nl
\enddata
\tablenotetext{a}{When only m=1 patterns are indicated, actual evolution
is apparently an accretion induced transient star/disk oscillation 
(see figure \ref{hiq-trans}) rather than a spiral arm.  NR (not 
resolved): for low stability disks, assignment of specific spiral arm
patterns loses meaning due to their rapid breakup.}

\tablecomments{Three series of runs are represented in this table. 
The first letter in each name is `s', signifying an SPH simulation.
The second, is either `c' or `q', signifying constant or varying \qmin,
and the third letter signifies that the simulation is a member of
a high (h), low (l) or varying (v) disk mass series.  Ascending
numerical order in each series refers to successive values of either
disk mass or \qmin, for each series.}
\end{deluxetable}

%% file: table-ppm.tex
\singlespace
\begin{deluxetable}{lcccccc}
\tablecolumns{7}
\tablecaption{\label{ppm-tabl} Disk Parameters for PPM Simulations}
\tablehead{
\colhead{Name}  & \colhead{Grid} & \colhead{\mrat} &
\colhead{\qmin}  & \colhead{End Time} & 
\colhead{Dominant\tablenotemark{a} } & \colhead{Outer} 
\\
\colhead{}  & \colhead{Res.} & \colhead{} &
\colhead{}  & \colhead{($T_D$=1)} & 
\colhead{Spiral Patterns}  & \colhead{Boundary}  }

\startdata
pcm1 &  64$\times$102 & 0.1 & 1.5 & 5.8  & NR    &  Refl.    \nl
pcm2 &  64$\times$102 & 0.2 & 1.5 & 5.0  & 2-4   &  Refl.    \nl
pcm3 &  64$\times$102 & 0.4 & 1.5 & 4.0  & 1-3   &  Refl.    \nl
pcm4 &  64$\times$102 & 0.6 & 1.5 & 3.75 & 1-3   &  Refl.    \nl
pcm5 &  64$\times$102 & 0.8 & 1.5 & 3.0  & 1-3   &  Refl.    \nl
pcm6 &  64$\times$102 & 1.0 & 1.5 & 3.0  & 1-3   &  Refl.    \nl
pch2 & 100$\times$152 & 0.2 & 1.5 & 5.0  & 2-4   &  Refl.    \nl
pch6 & 100$\times$152 & 1.0 & 1.5 & 3.6  & 1-3   &  Refl.    \nl
pqm1 &  64$\times$102 & 0.8 & 1.1 & 1.8  & 1-2   &  Refl.    \nl
pqm2 &  64$\times$102 & 0.8 & 1.3 & 2.6  & 1-2   &  Refl.    \nl
pqm3 &  64$\times$102 & 0.8 & 1.4 & 3.0  & 1-2   &  Refl.    \nl
pqm4 &  64$\times$102 & 0.8 & 1.7 & 3.0  & 1-2   &  Refl.    \nl
pqm5 &  64$\times$102 & 0.8 & 2.0 & 2.0  & 1     &  Refl.    \nl
pci2 &  64$\times$96  & 0.2 & 1.5 & 3.8  & 3-4   &  Infall   \nl
pci3 &  64$\times$96  & 0.5 & 1.5 & 2.1  & 1-3   &  Infall   \nl
pci4 &  64$\times$96  & 0.6 & 1.5 & 2.0  & 1,3   &  Infall   \nl
pci6 &  64$\times$96  & 1.0 & 1.5 & 1.6  & 1-3   &  Infall   \nl
pcl1 &  44$\times$64  & 0.1 & 1.5 & 5.6  & NR    &  Infall   \nl
pcl2 &  44$\times$64  & 0.3 & 1.5 & 4.2  & 1-3   &  Infall   \nl
pcl3 &  44$\times$64  & 0.4 & 1.5 & 4.2  & 1-3   &  Infall   \nl
pcl4 &  44$\times$64  & 0.5 & 1.5 & 2.8  & 1-2   &  Infall   \nl
pcl5 &  44$\times$64  & 0.7 & 1.5 & 2.8  & 1-2   &  Infall   \nl
pcl6 &  44$\times$64  & 1.0 & 1.5 & 2.0  & 1-2   &  Infall   \nl
\enddata
\tablenotetext{a}{NR: not resolved. For some low mass disks,
distinct spiral patterns are not possible to distinguish.}

\tablecomments{Each of these PPM runs begins with `p' to distinguish
it from SPH series. The second letter is `c' or `q' signifying a
constant or varying \qmin~value for each disk in the series.
The third letter implies a {\it l}ow, {\it m}oderate or {\it h}igh
resolution simulation. Moderate resolution infall boundary
simulations are distinquished from reflecting boundary simulations
using an `i' in place of `m'.  Numbers are successive values of disk 
mass or \qmin in each series of runs.}
\end{deluxetable}

%% file: table-cmp.tex
\singlespace
\begin{deluxetable}{lccccc}
\tablecolumns{6}
\tablecaption{\label{cmp-tabl} Tori and Disk Results in SPH and PPM}
\tablehead{
\colhead{Initial}   & \colhead{Hydro}      & \colhead{Linear} &
\colhead{Reason/}    & \colhead{Non-linear} & \colhead{Reason/}
\\
\colhead{Density}   & \colhead{Method} & \colhead{Regime} &
\colhead{Result}    & \colhead{Regime} & \colhead{Result}
\\
\colhead{Structure} & \colhead{}       & \colhead{}       &
\colhead{}          & \colhead{}       & \colhead{}
}

\startdata
Disk         & PPM & fails        &  inner       & inaccessible  & \nodata            \nl
(eq. \ref{denslaw})& &            & boundary     &               &                    \nl
             & SPH & inaccessible & short time   & succeeds      & spiral arm         \nl
             &     &              & baseline     &               & formation/collapse \nl

Disk w/ Central & PPM & succeeds  & spiral arm   & short         & boundary  \nl
Hole (eq. \ref{denslaw-a}) &  &   & growth       & duration only & influence \nl
             & SPH & inaccessible & short time   & \nodata       & \nodata   \nl
             &     &              & baseline     &               &           \nl

Torus        & PPM & succeeds     & spiral arm   & succeeds     & spiral arm \nl
(eq. \ref{torus-law})& &          & growth       &              & collapse   \nl
	     & SPH & partial      & spiral arm   & succeeds     & spiral arm \nl
             &     & success      & growth       &              & collapse   \nl
\enddata
\end{deluxetable}

%% file: figures.tex
\begin{figure}
\plotfiddle{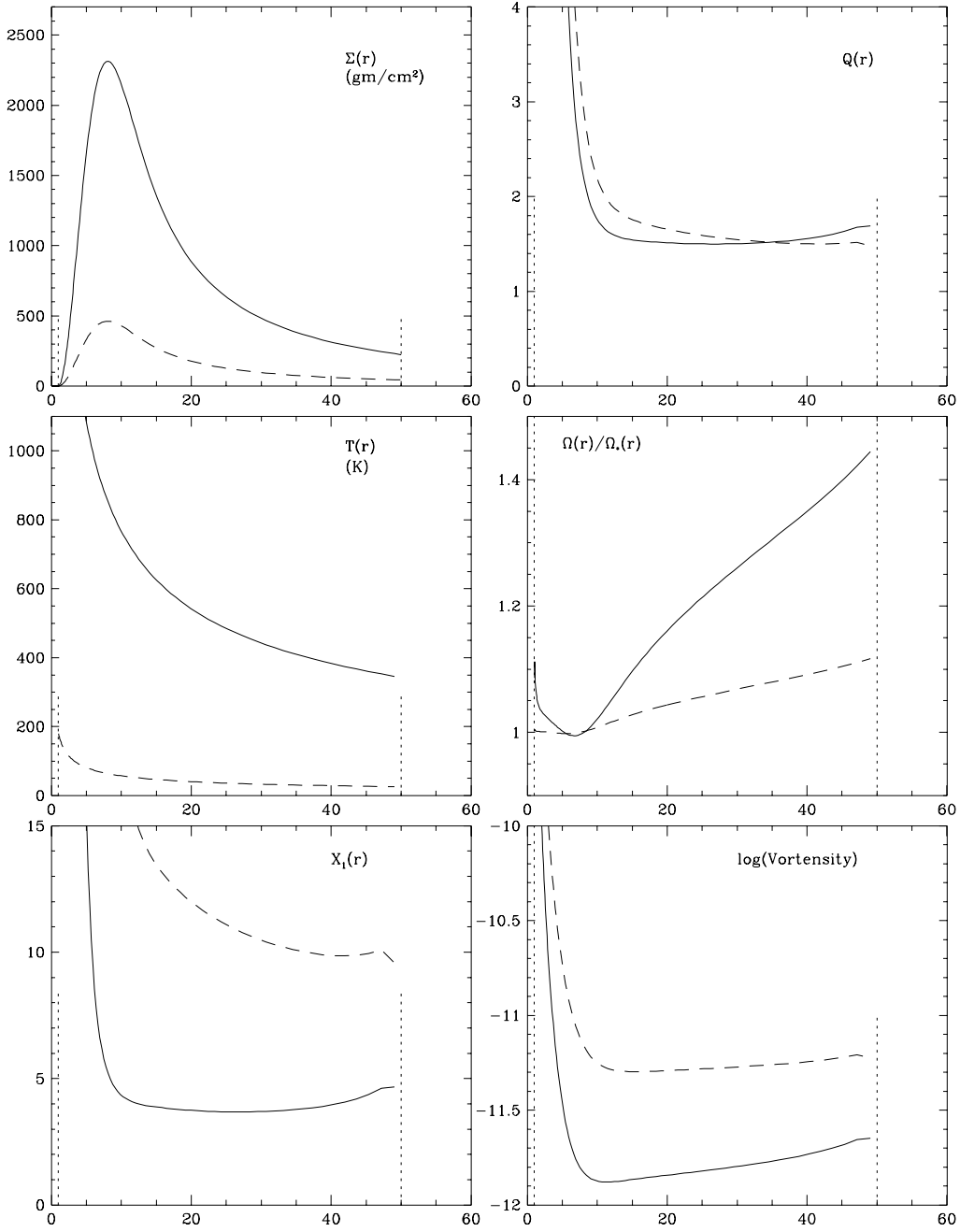}{6.75in}{0}{75}{75}{-240}{-50}
\caption{\label{dinit-ppm}
A summary of the initial conditions for low (dashed) and high (solid)
disk mass PPM simulations (simulations {\it pch2} and {\it pch6}).
The six panels show surface density $\Sigma$, Toomre $Q$, temperature
$T$, the ratio of the rotation period at radius $r$ with the Keplerian
value. We define $\Omega_*(r)$ in the middle right panel as
$\Omega_*(r) = \sqrt{ {{GM_*}/{r^3}}}$. In the lower left
panel, we show the value of the SWING $X$ parameter for the $m=1$
pattern.  Higher order patterns ($m>1$) are smaller by a factor $1/m$
than the value shown. In the lower right panel, we show the value
of the vortensity at each radius.}
\end{figure}

\clearpage

\begin{figure}
\plotfiddle{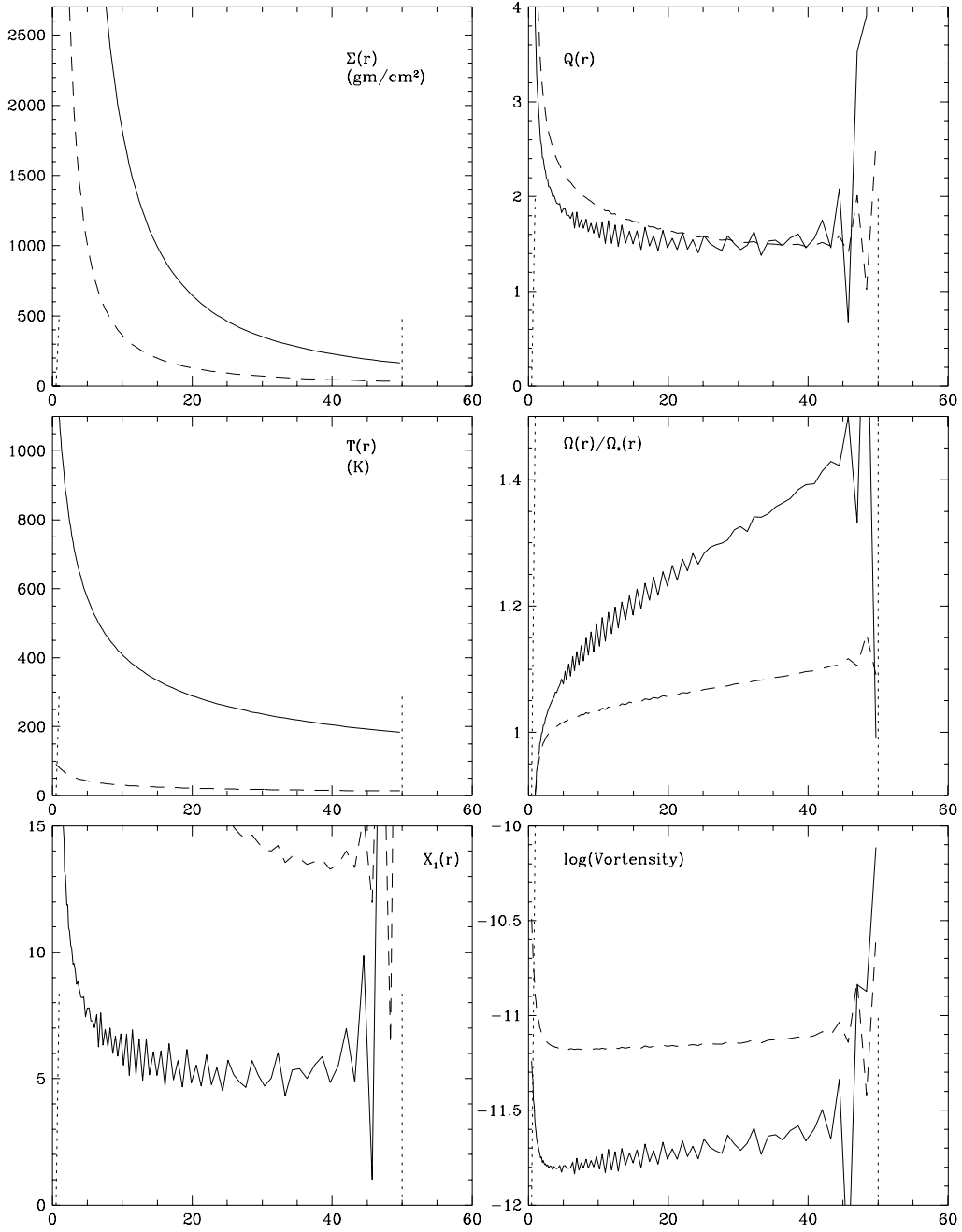}{6.75in}{0}{75}{75}{-240}{-50}
\caption{\label{dinit-sph}
A summary of the initial conditions for low (dashed) and high (solid) 
disk mass SPH simulations (simulations {\it scv2} and {\it scv6}). This 
figure shows the same parameters as shown in figure \ref{dinit-ppm}.}
\end{figure}

\clearpage

\begin{figure}
\plotfiddle{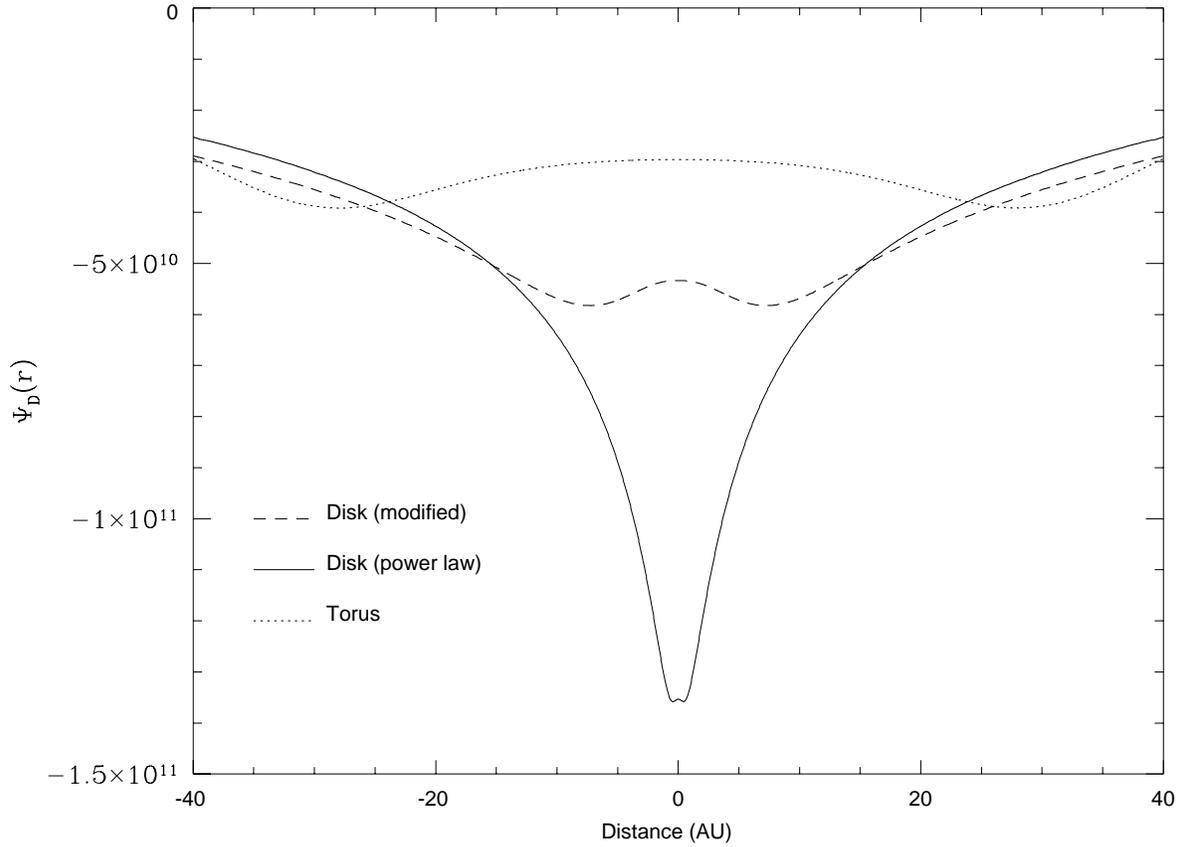}{5in}{-90}{60}{60}{-230}{400}
\caption{\label{gravpot}
The initial gravitational potential due to the disks and tori we 
study. We show a slice through the origin where the star is initially
located. The solid curve represents the gravitational potential
due to the pure power law as given by eq.  [\ref{denslaw}].  The
dashed curve is that due to the modified power law of eq.  
[\ref{denslaw-a}], while the dotted curve is that due to 
a torus as defined in section \ref{sphppm}. The mass in each disk
or torus is \mrat~=0.2.  Each system produces a gravitational 
potential hump at the origin which seeds the growth of $m=1$
disturbances in the disk or torus.}
\end{figure}

\clearpage

\begin{figure}
\plotfiddle{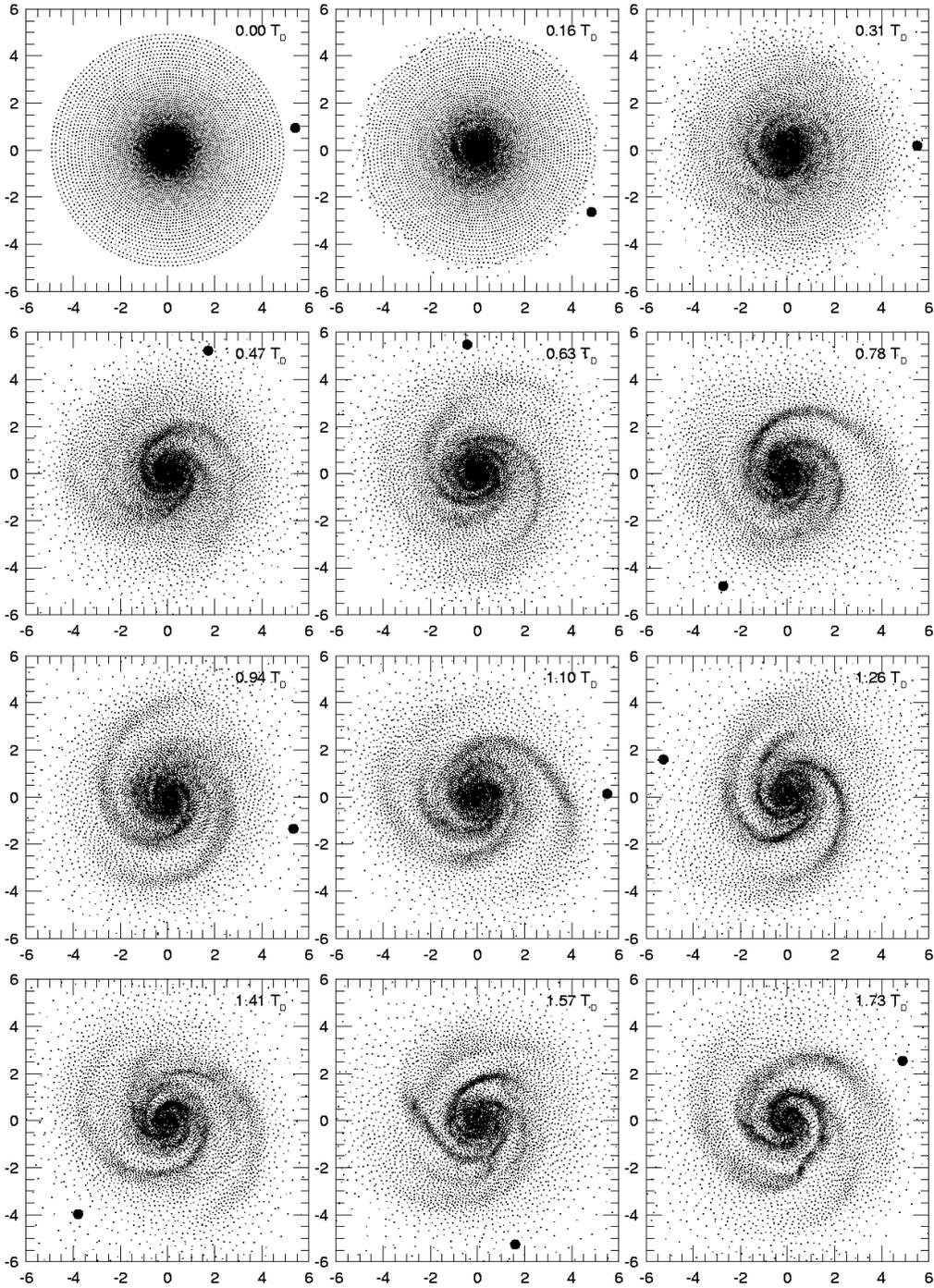}{6.79in}{0}{72}{72}{-230}{-20}
\caption{\label{sph-himas}
A time series mosaic of SPH particle positions for a disk of mass
\mrat~$=1.0$ and \qmin~$=1.5$ (simulation {\it scv6}).
Note the strong variation of spiral structure over time. Length units
are defined as 1=10 AU and time in units of the disk orbit period $T_D$.
The large, solid dot is the angular position of the star projected
out to a distance of 55 AU, just outside the outer disk edge.
The final image in this mosaic shows the beginning stages of clump
formation as a clump begins to form in the disk at about an azimuth
angle of 5 o'clock and radius of $r=20$ AU.  A second clump which initially
formed in the other spiral arm is present but difficult to distinguish
in the image at 3 o'clock and $r\sim7$ AU.}
\end{figure}

\clearpage

\begin{figure}
\plotfiddle{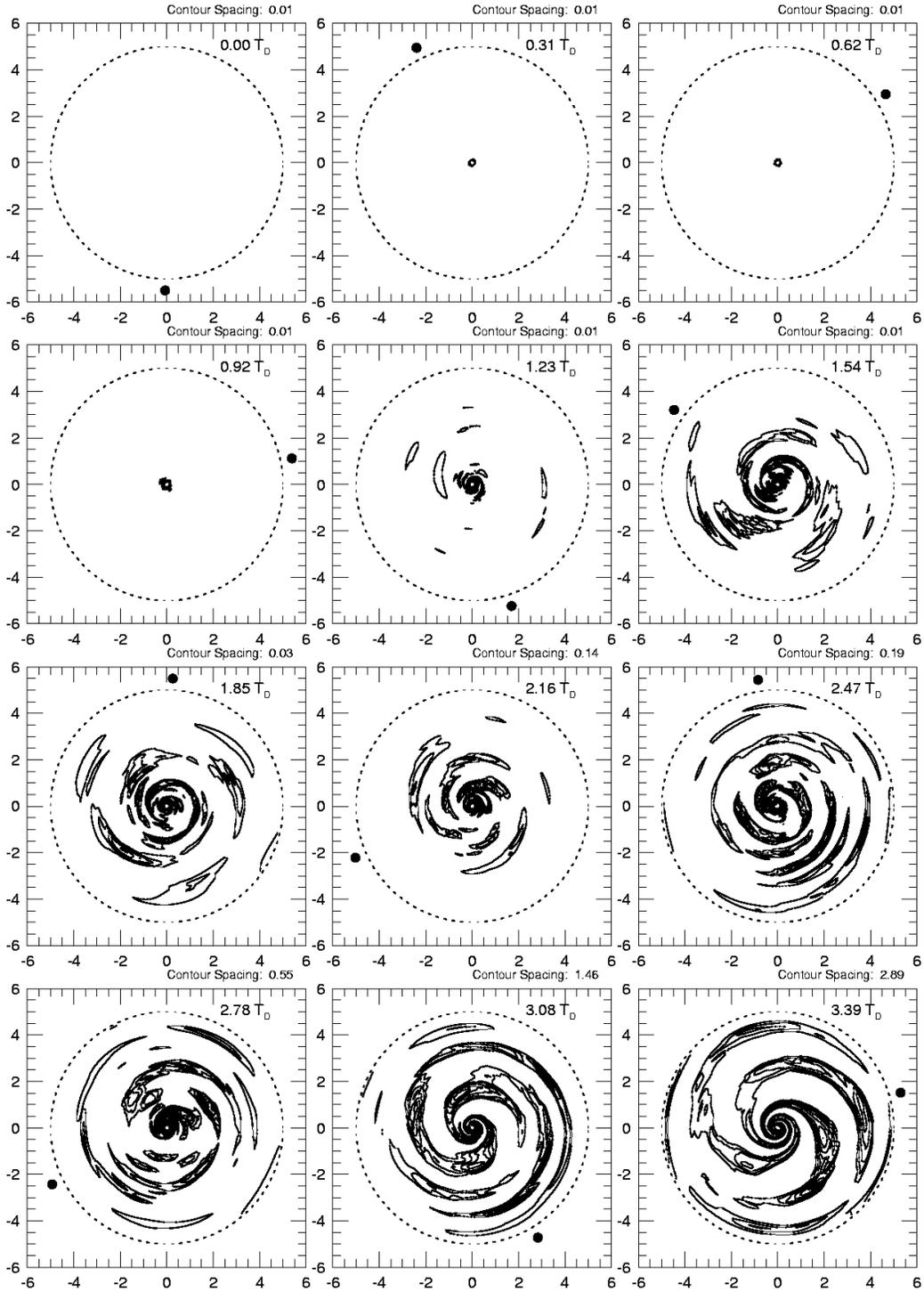}{7.15in}{0}{75}{75}{-230}{-15}
\caption{\label{ppm-himas}
Time series of density variation in the disk for a PPM simulation
(simulation {\it pch6}). with the same initial conditions as figure
\ref{sph-himas}. Only positive density variations are plotted and the
maximum contour is derived only from deviations at radii larger than
$\sim$7 AU. Contours are linearly spaced and set to a minimum of 
.01\% variation per contour line. Larger variations are implemented
as the instability grows. The contour spacing is denoted in the
upper right corner of each frame.}
\end{figure}

\clearpage

\begin{figure}
\plotfiddle{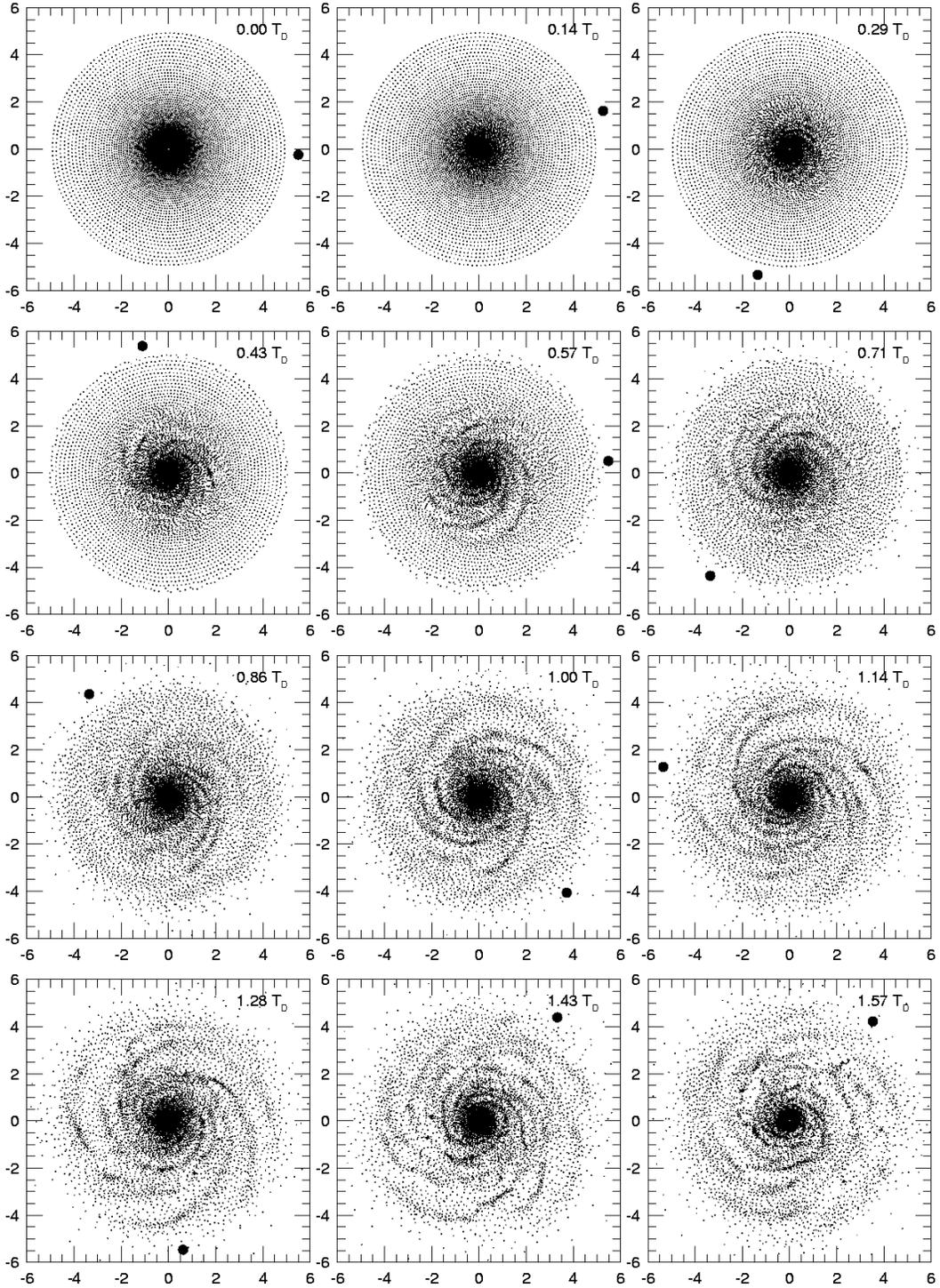}{7.25in}{0}{75}{75}{-230}{-10}
\caption{\label{sph-lomas}
Evolution of a disk with \mrat~$=0.2$ and with initial 
\qmin~$=1.5$ (simulation {\it scv2}).  Note the production of long
filamentary spiral arms and the production of multiple clumps at later 
times.}
\end{figure}

\clearpage

\begin{figure}
\plotfiddle{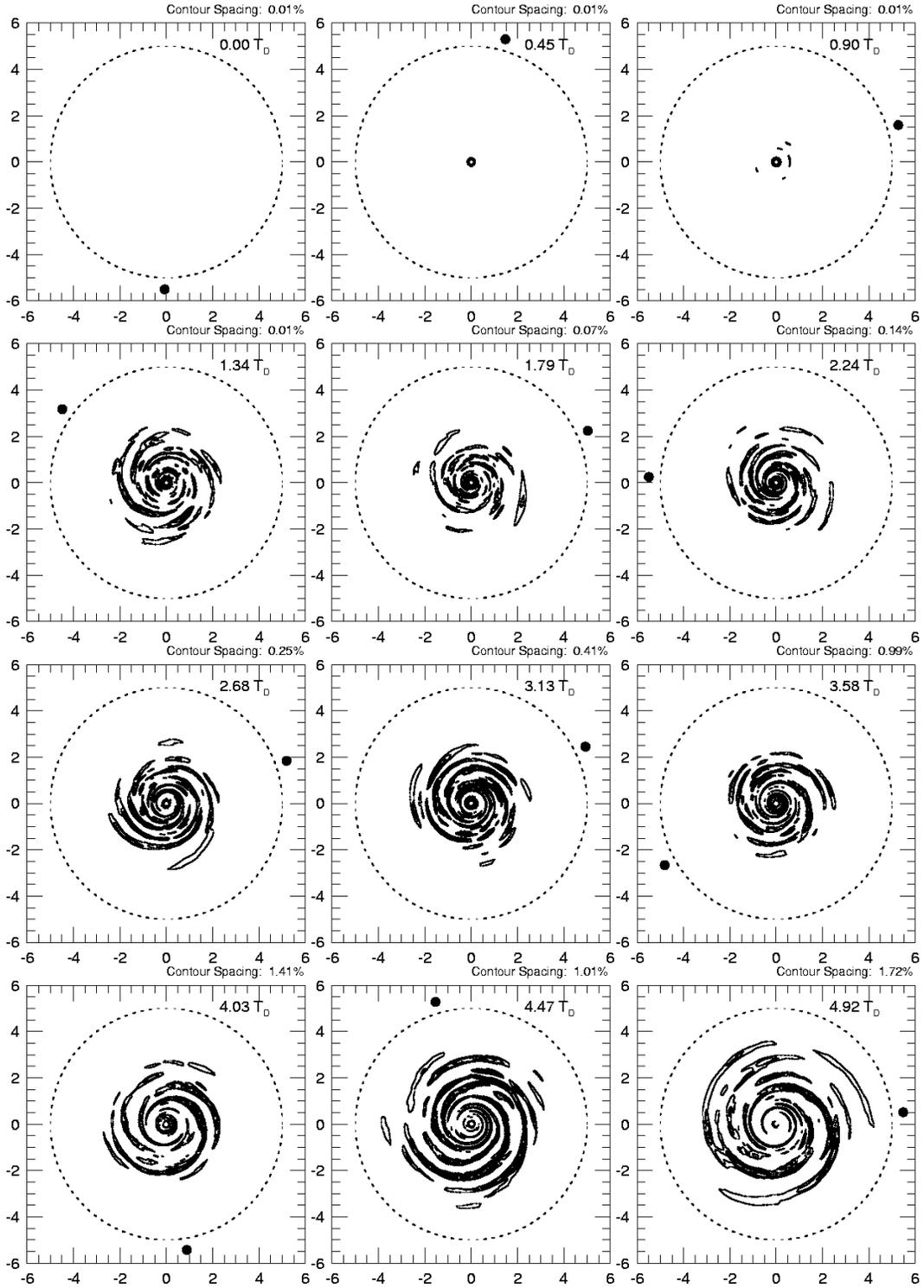}{7.25in}{0}{77}{77}{-230}{-15}
\caption{\label{ppm-lomas}
The same initial conditions as figure \ref{sph-lomas} with the PPM code
(simulation {\it pch2}).  A much longer evolution than figure 
\ref{sph-lomas} is possible here due to the low initial noise of PPM.}
\end{figure}

\clearpage

\begin{figure}
\plotfiddle{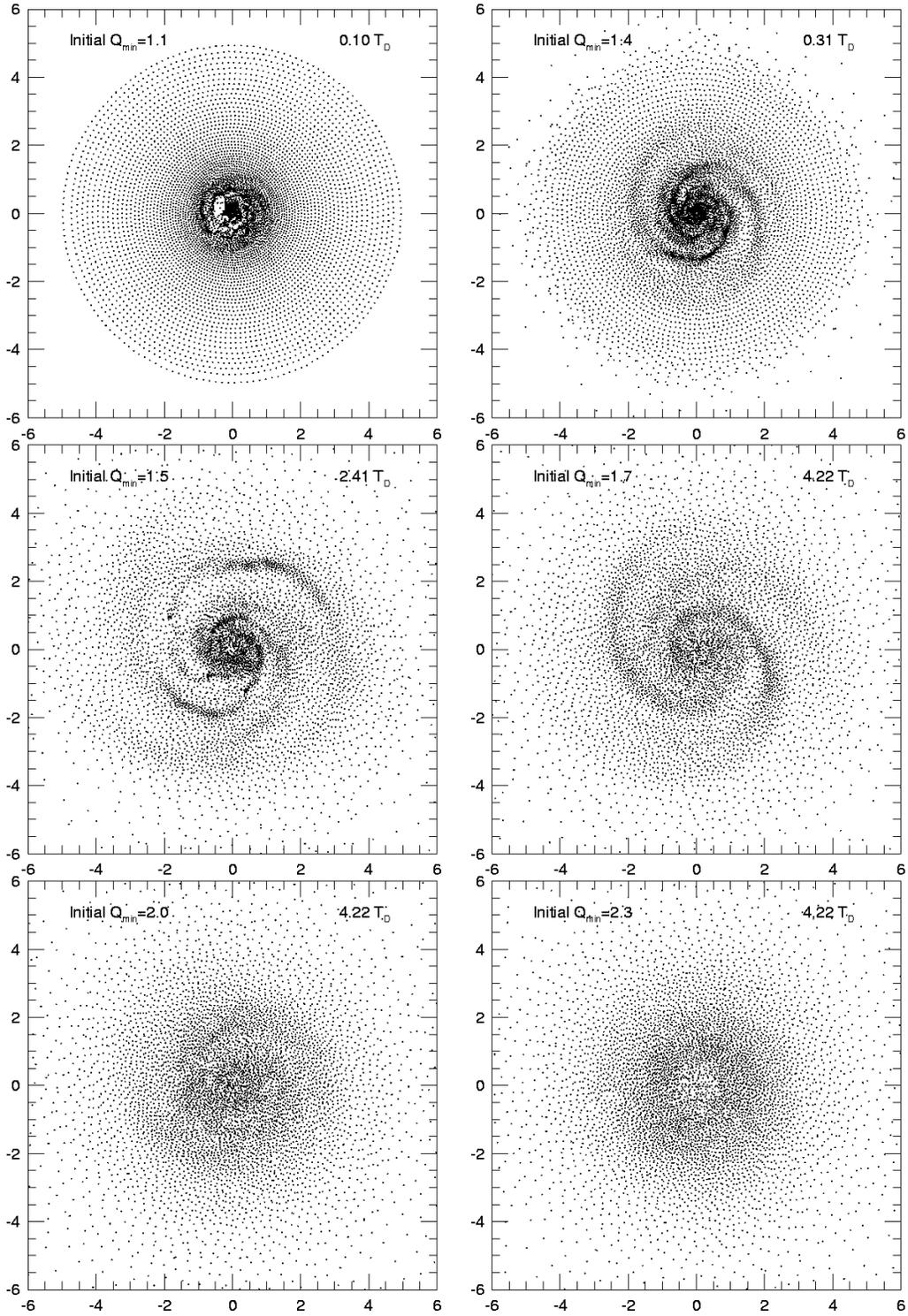}{7.25in}{0}{77}{77}{-240}{-15}
\caption{\label{sph-qvar}
Late time snapshots of a series of disk simulations using
our SPH code. Each disk has the same disk mass of \mrat~$=0.8$ but 
varying \qmin (simulations {\it sqh1, -3,-4, -5}, and {\it -6}, as well
as {\it scv5} are shown).}
\end{figure}

\clearpage

\begin{figure}
\plotfiddle{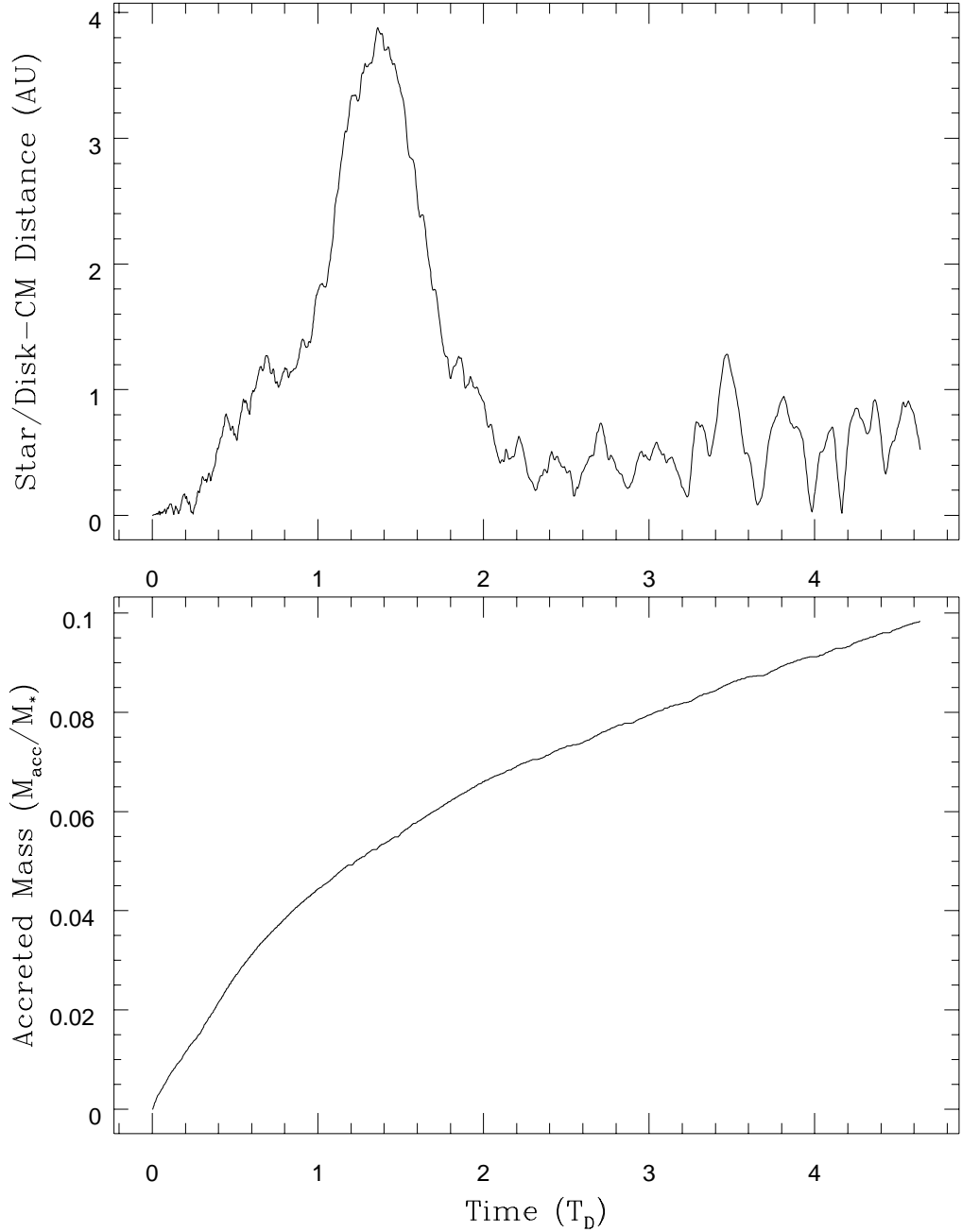}{6.5in}{0}{70}{70}{-200}{-30}
\caption{\label{hiq-trans}
The distance between the star and the disk center of mass is
shown as a function of time in the top panel here, while the mass
accreted by the star is shown in the second. The simulation these
data are taken from is {\it sqh6}, which begins with \mrat~$=0.8$
and an initial minimum $Q$ value of 2.3.  With the units assumed for
our systems, the mass accretion rate is near $8\times 10^{-5} M_\odot$/yr
at its maximum.  When accretion begins to drain the inner disk
matter and the rate falls sufficiently (in this simulation, to
$\sim 3\times 10^{-5} M_\odot$/yr), the star falls to the center
of the system and returns much of its temporary increase in angular
momentum to the disk.  }
\end{figure}

\clearpage

\begin{figure}
\plotfiddle{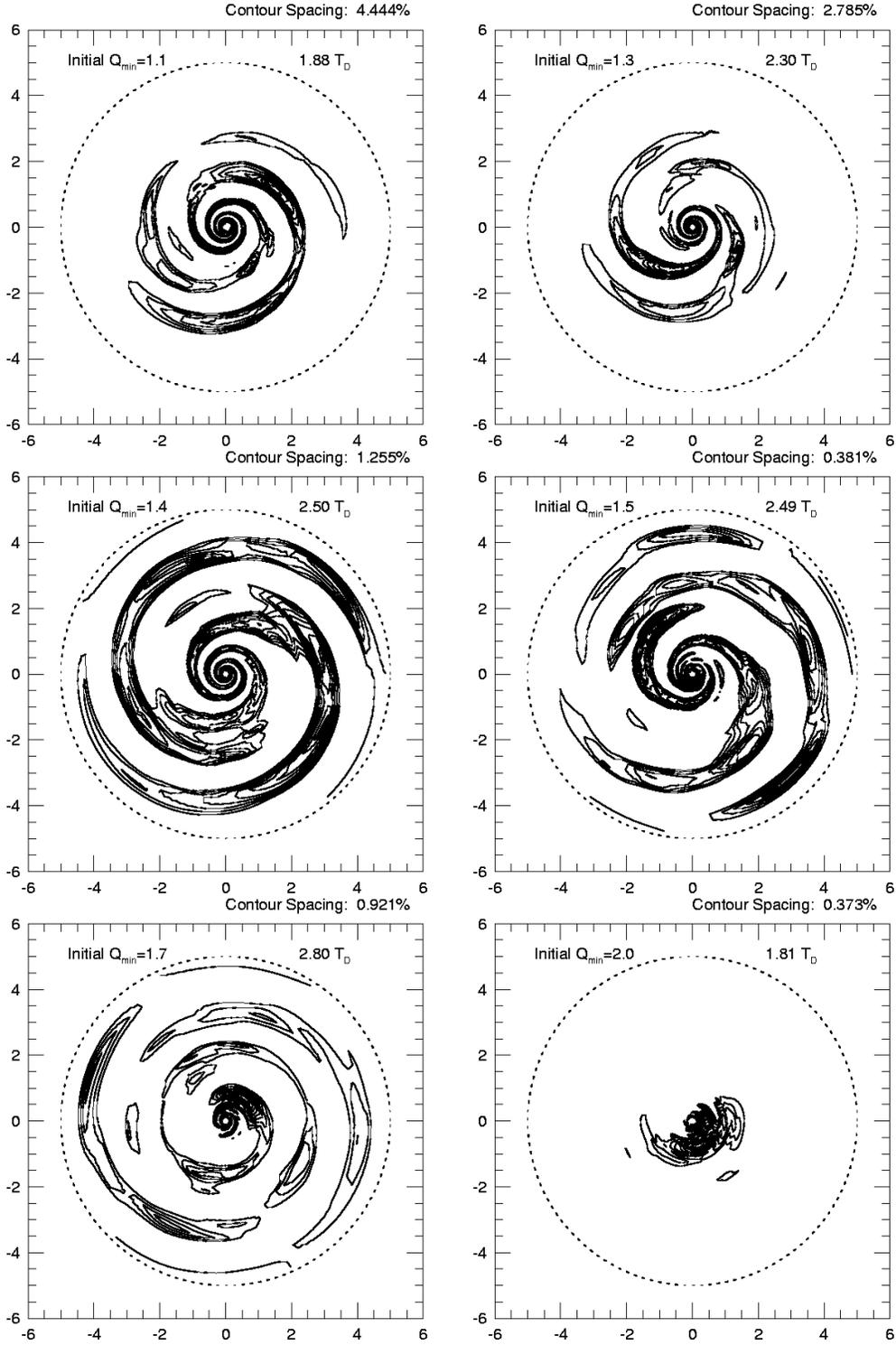}{7.25in}{0}{77}{77}{-240}{-15}
\caption{\label{ppm-qvar}
Late time snapshots of a series of disk simulations using
our PPM code. Each disk has the same disk mass of \mrat~$=0.8$ but 
varying \qmin (simulations {\it pqm1-5} as well as {\it pcm5} are
shown).}
\end{figure}

\clearpage

\begin{figure}
\plotfiddle{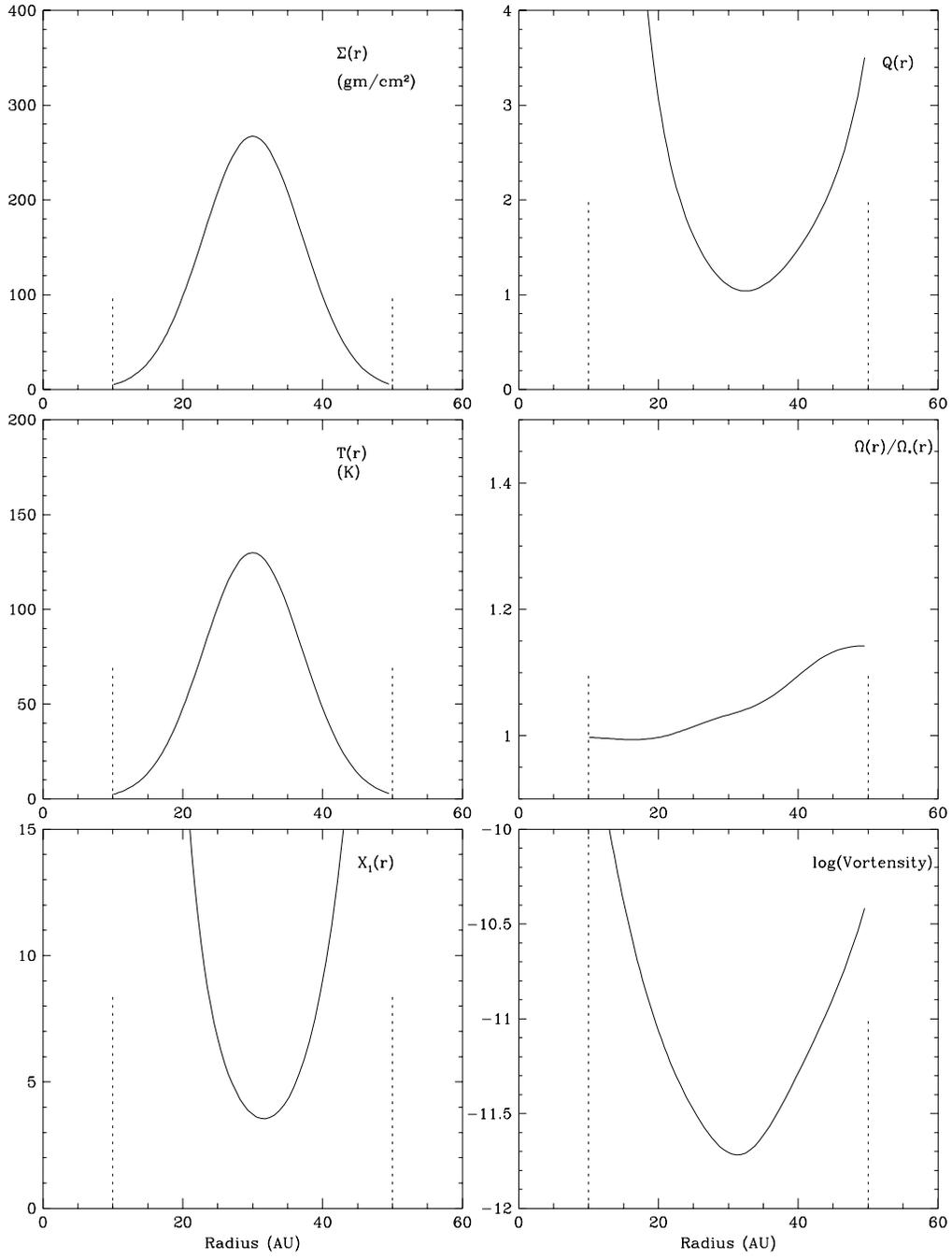}{6.75in}{0}{75}{75}{-240}{-50}
\caption{\label{torus-init}
Initial conditions for torus simulations.  Each frame contains the
same variable as in the corresponding frames in figures \ref{dinit-ppm}
and \ref{dinit-sph}.}
\end{figure}

\clearpage

\begin{figure}
\plotfiddle{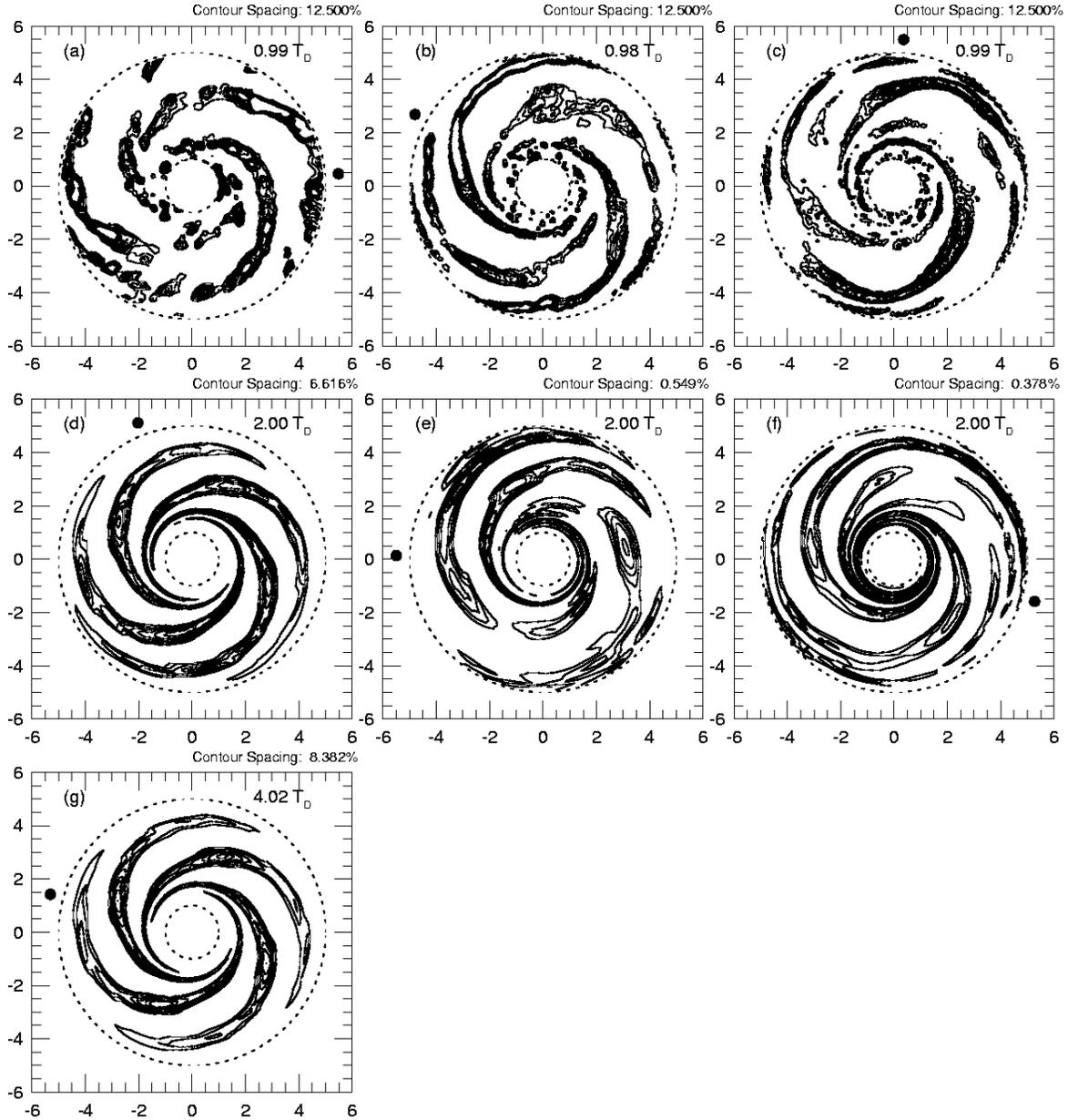}{6.65in}{0}{75}{75}{-240}{-20}
\caption{\label{tor-cmp}
Late time snapshots of a torus with identical initial
conditions using (a-c) SPH with $\sim$7000, 14000, and 28000 particles
respectively. (d-f) PPM with 10$^{-3}$ amplitude random initial noise
at three grid resolutions: 40$\times$150, 60$\times$225 and
80$\times$300 and (g) PPM simulation with low initial noise (10$^{-8}$)
at 40$\times$150 grid resolution. For comparison purposes, the SPH runs
are mapped onto a grid identical to that used for the corresponding
PPM runs.}
\end{figure}

\clearpage

\begin{figure}
\plotfiddle{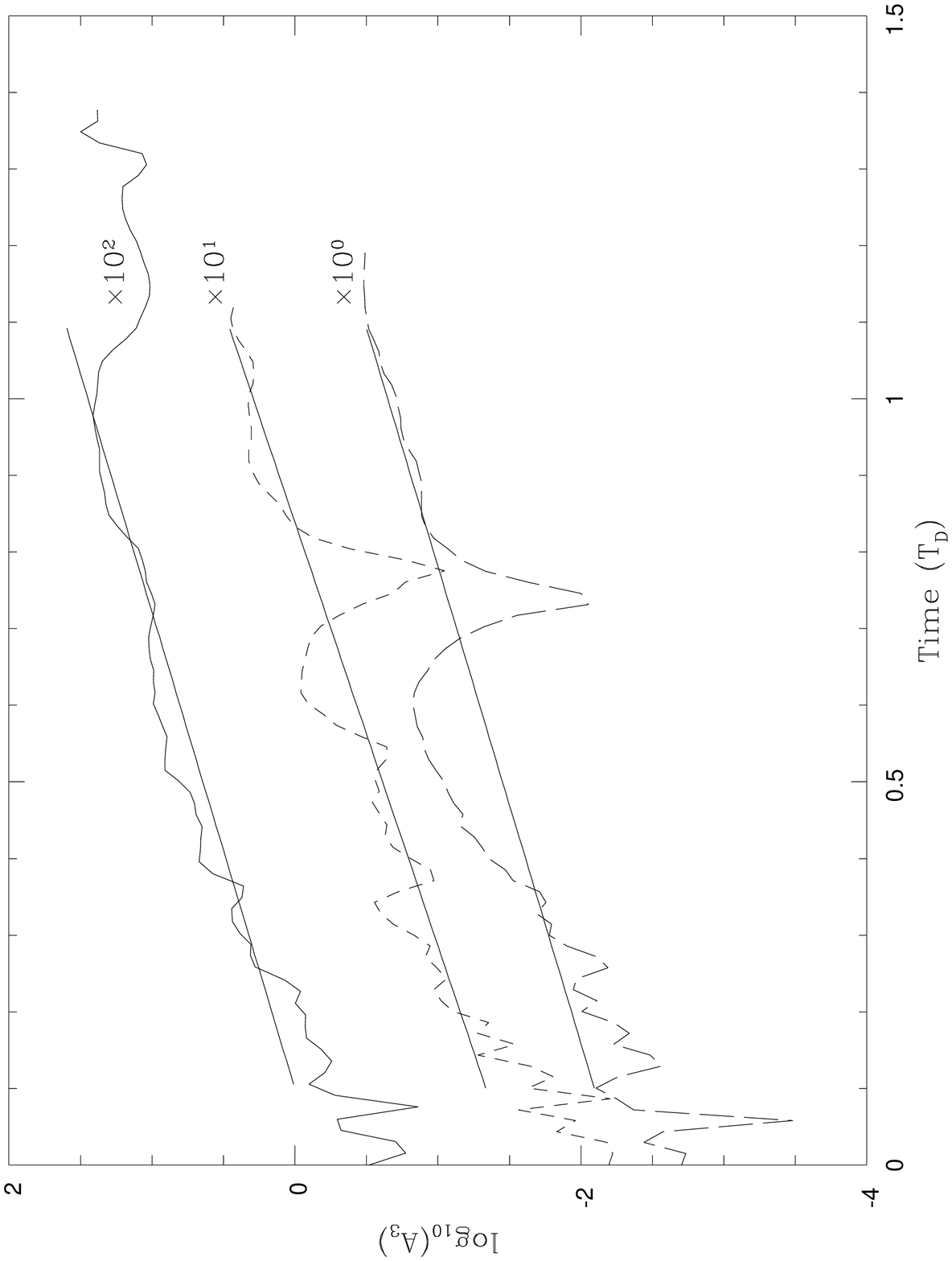}{3.02in}{-90}{45}{45}{-180}{265}
\plotfiddle{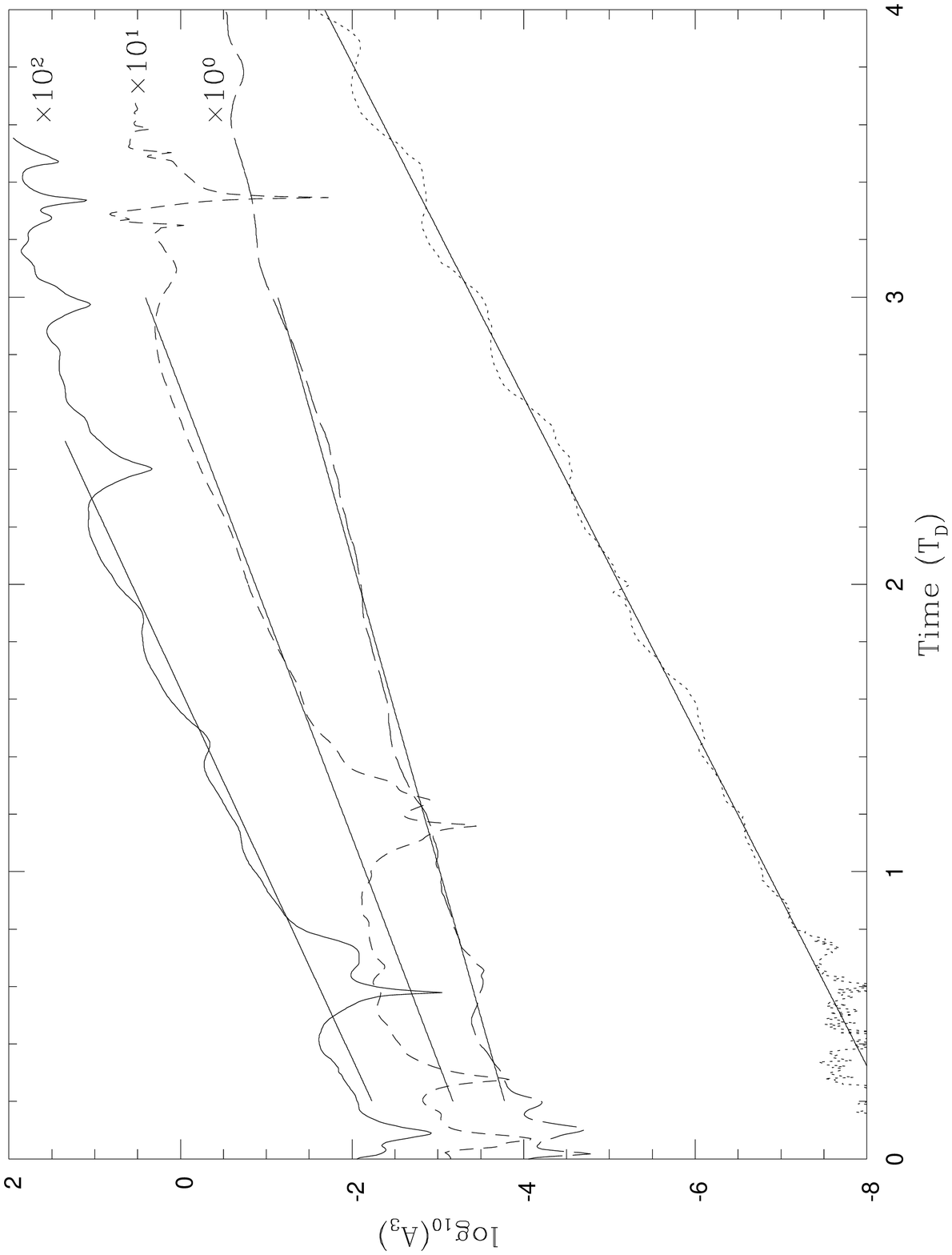}{3.02in}{-90}{45}{45}{-180}{250}
\caption{\label{torm3_30}
Amplitudes and linear best fits for the $m=3$ pattern at the center of 
the torus ($R=$30 AU) for different resolution SPH and PPM simulations.
The top panel shows SPH simulations. The lowest resolution ($\sim 7000$ 
particles) is denoted with a solid curve while double resolution ($\sim
14000$ particles) is denoted with a short dashed curve and the highest 
resolution ($\sim 28000$ particles) is shown with a long dashed curve.  
Each of the fits are shown as solid lines. Bottom panel: PPM simulations
with the two lowest resolution runs denoted by a solid and dotted line for
the 10$^{-3}$ and 10$^{-8}$ amplitude initial noise runs respectively.
The short dashed curve represents the middle resolution and the long 
dashed line represents the highest resolution run. Solid lines denote
the best fit curves for each of the runs and displayed only for the
times for which the fit was derived. Each of the SPH runs and the PPM 
runs with 10$^{-3}$ noise are artificially multiplied by a factor of
1, 10 or 100 in order to distinguish between the different runs on the
plots.}
\end{figure}

\clearpage

\begin{figure}
\plotfiddle{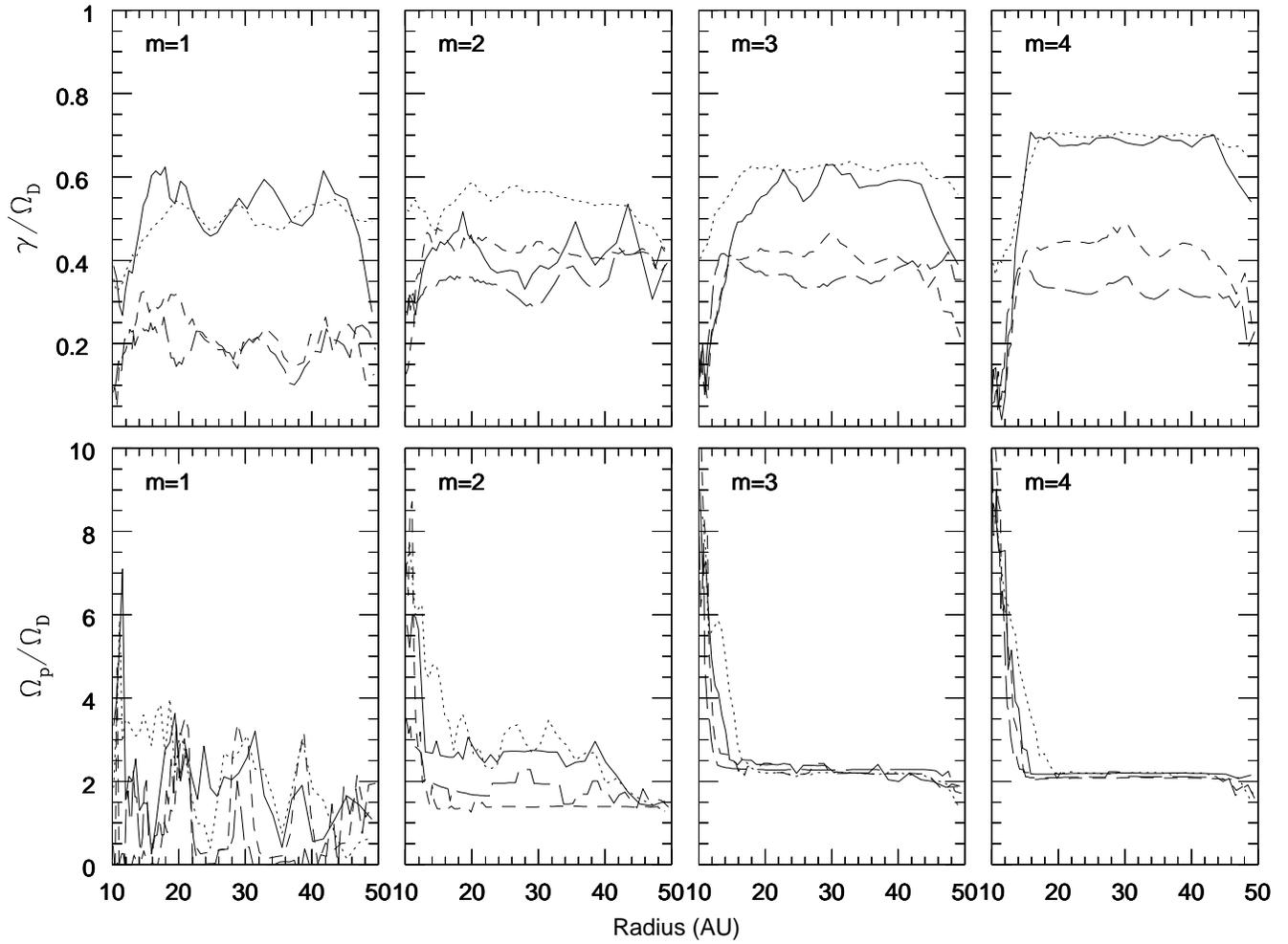}{5.5in}{-90}{70}{70}{-280}{500}
\caption{\label{ppmpatgrw1-4}
Growth rates and pattern speeds for the $m=1-4$ patterns
derived from PPM simulations. The increase in the pattern
speed at the inner torus edge probably represents a boundary 
influence and we do not consider it to be significant. Each curve
uses the same representation as in figure \ref{torm3_30} to denote
low, moderate and high resolution runs. }
\end{figure}

\clearpage

\begin{figure}
\plotfiddle{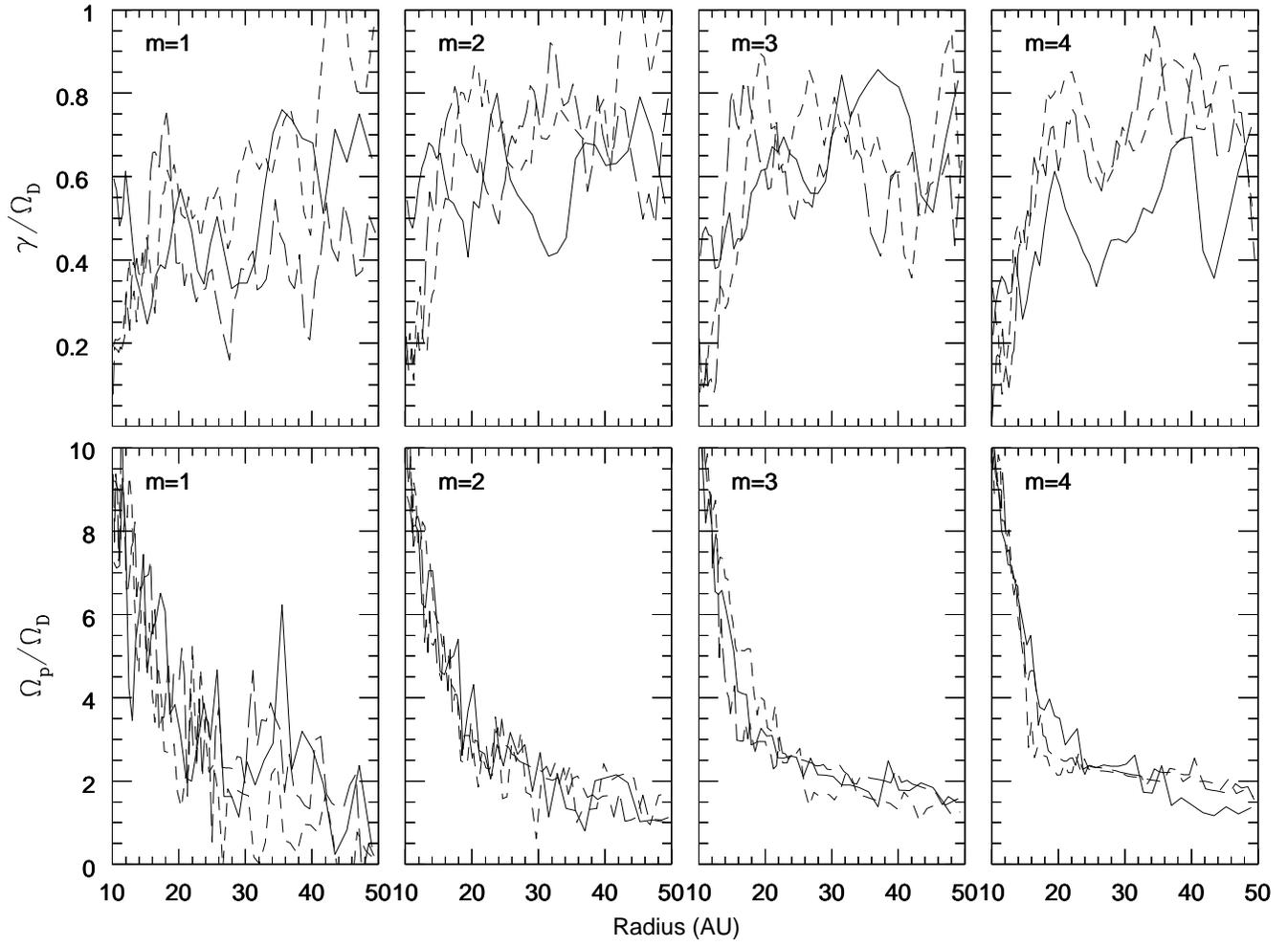}{5.5in}{-90}{70}{70}{-280}{540}
\caption{\label{sphpatgrw1-4}
Growth rates and pattern speeds for the $m=1-4$ patterns derived from 
the SPH simulations. Each curve uses the same representation as in
figure \ref{torm3_30} to denote low moderate and high resolution
runs. }
\end{figure}

\clearpage

\begin{figure}
\plotfiddle{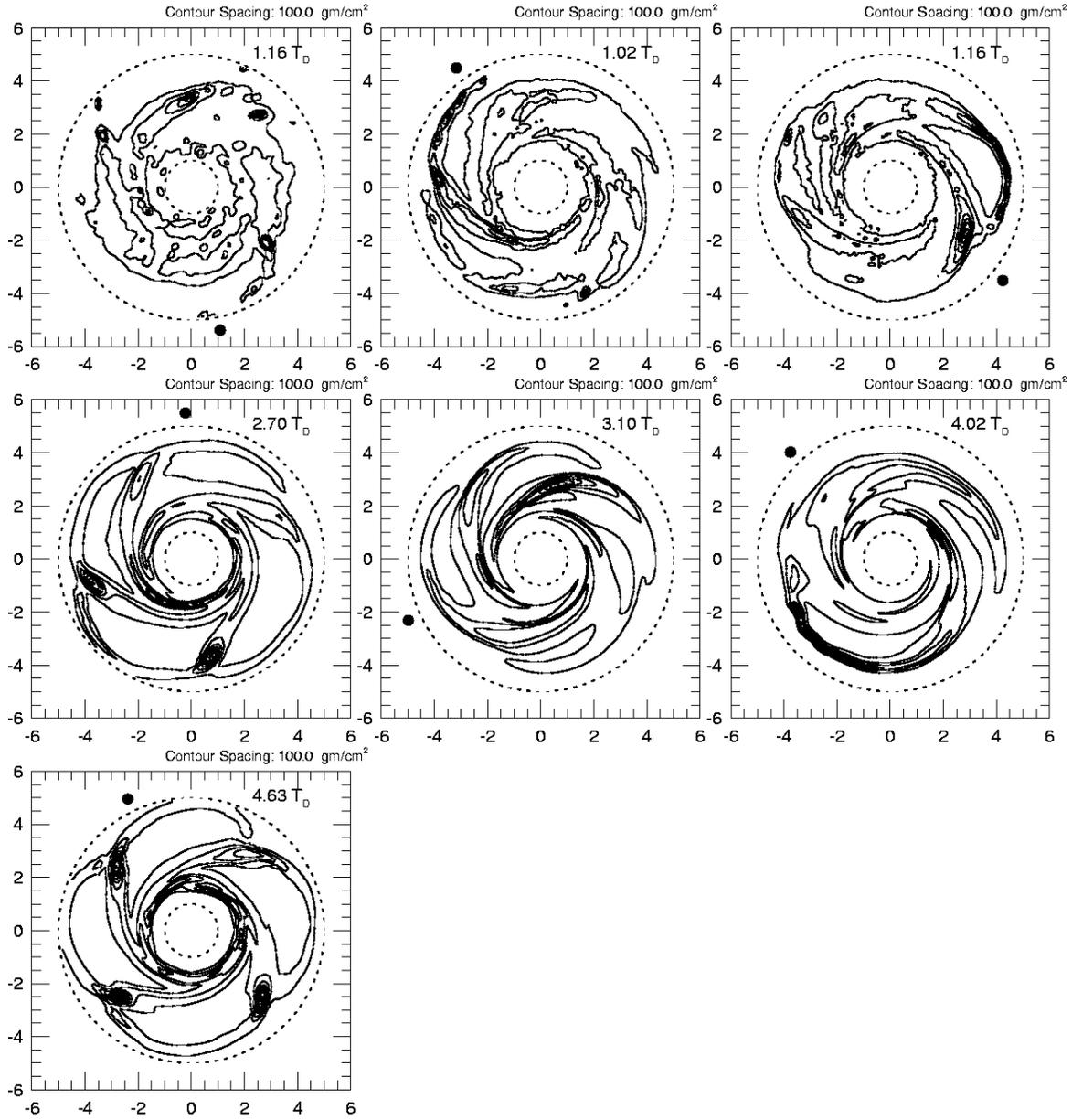}{6.6in}{0}{75}{75}{-240}{-20}
\caption{\label{tor-late}
Late time snap shots of the same simulations as figure \ref{tor-cmp}
above. Here we plot density rather than density variation to accentuate
collapse behavior. Contours units are gm/cm$^2$ and are linearly spaced
from 0 gm/cm$^2$ (not shown) upward with spacing between contours as
noted at the upper right of each frame. Because the collapse behavior
occurs at a somewhat different time for each of the runs, the plots
are not shown at the same time as any other plot. Rather, we show
the morphology shortly after collapse begins in each simulation,
at whatever time during the simulation that occurred. Each of the SPH
runs are mapped onto a grid identical to that used by the corresponding
PPM simulation. The dashed curves denoting the inner and outer grid
radii therefore have no meaning for these runs.}
\end{figure}

\clearpage

\begin{figure}
\plotfiddle{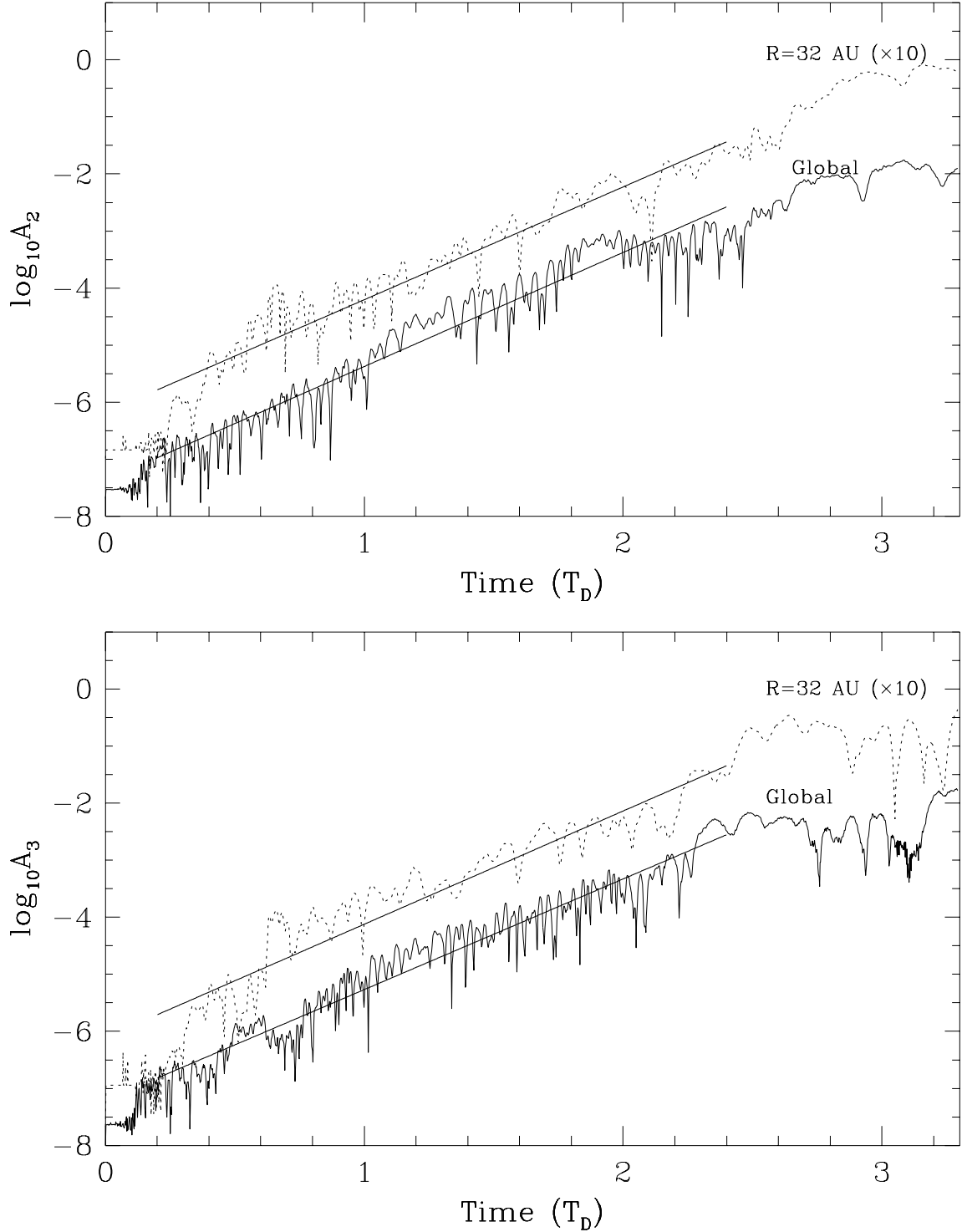}{7.5in}{0}{80}{80}{-250}{-25}
\caption{\label{m2and3hi}
The amplitudes and fits for the $m=2$ (top frame) and $m=3$ (bottom
frame) patterns derived from the simulation shown in figure 
\ref{ppm-himas}. The amplitude ($\times$~10) near the middle of
the power law portion of the disk as well as the globally integrated 
amplitudes for each pattern are shown. }
\end{figure}

\clearpage

\begin{figure}
\plotfiddle{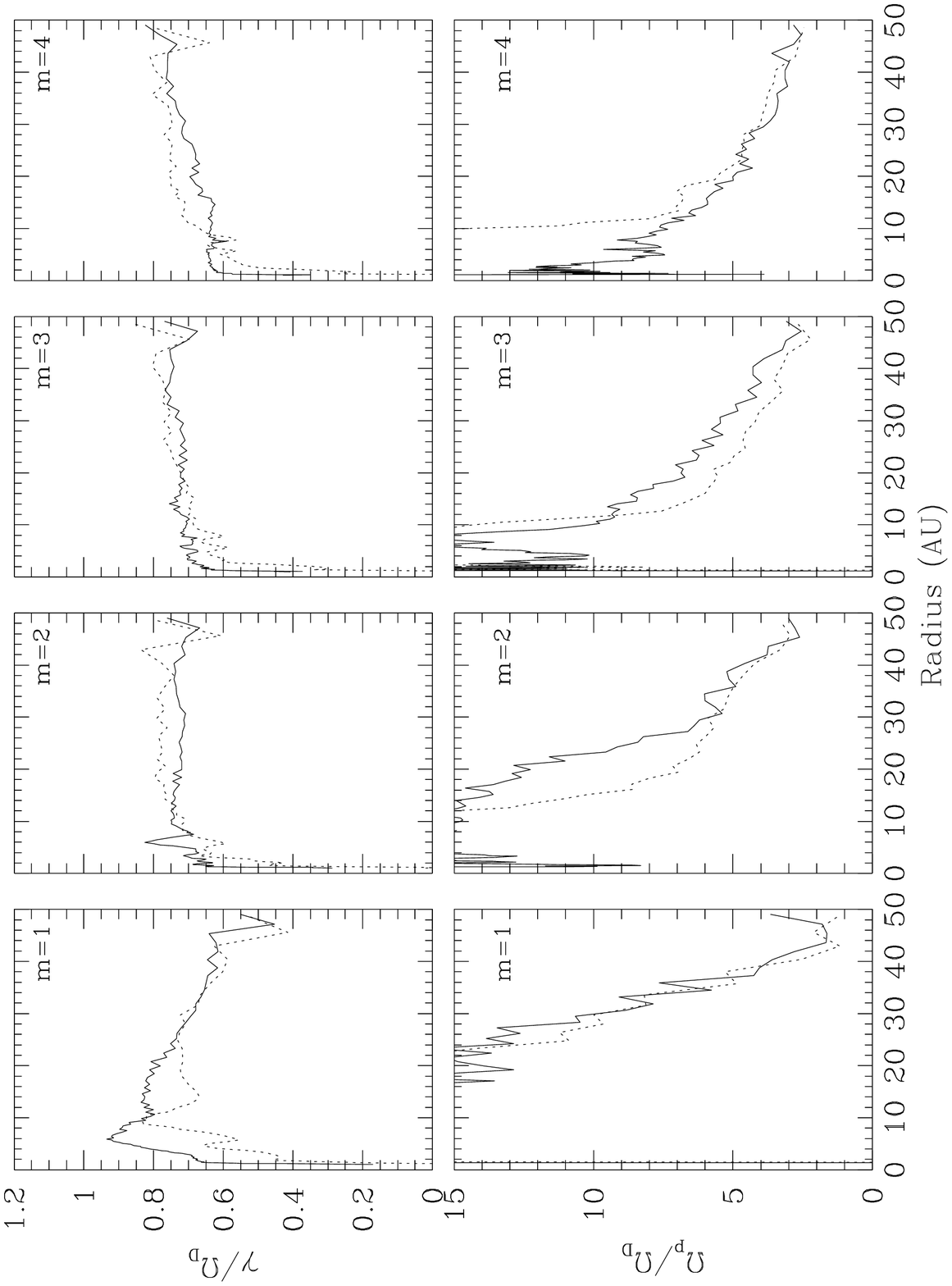}{5.5in}{-90}{70}{70}{-280}{500}
\caption{\label{patgrwm1-4hi}
The growth rates and pattern speeds for the $m=1$--4 spiral
arm patterns. The simulation from which these are derived is the same
as is shown in figure \ref{ppm-himas}. The solid lines represent the
moderate resolution simulation {\it pch6}  while the dotted lines 
represent results from the lower resolution simulation {\it pcm6}.}
\end{figure}

\clearpage

\begin{figure}
\plotfiddle{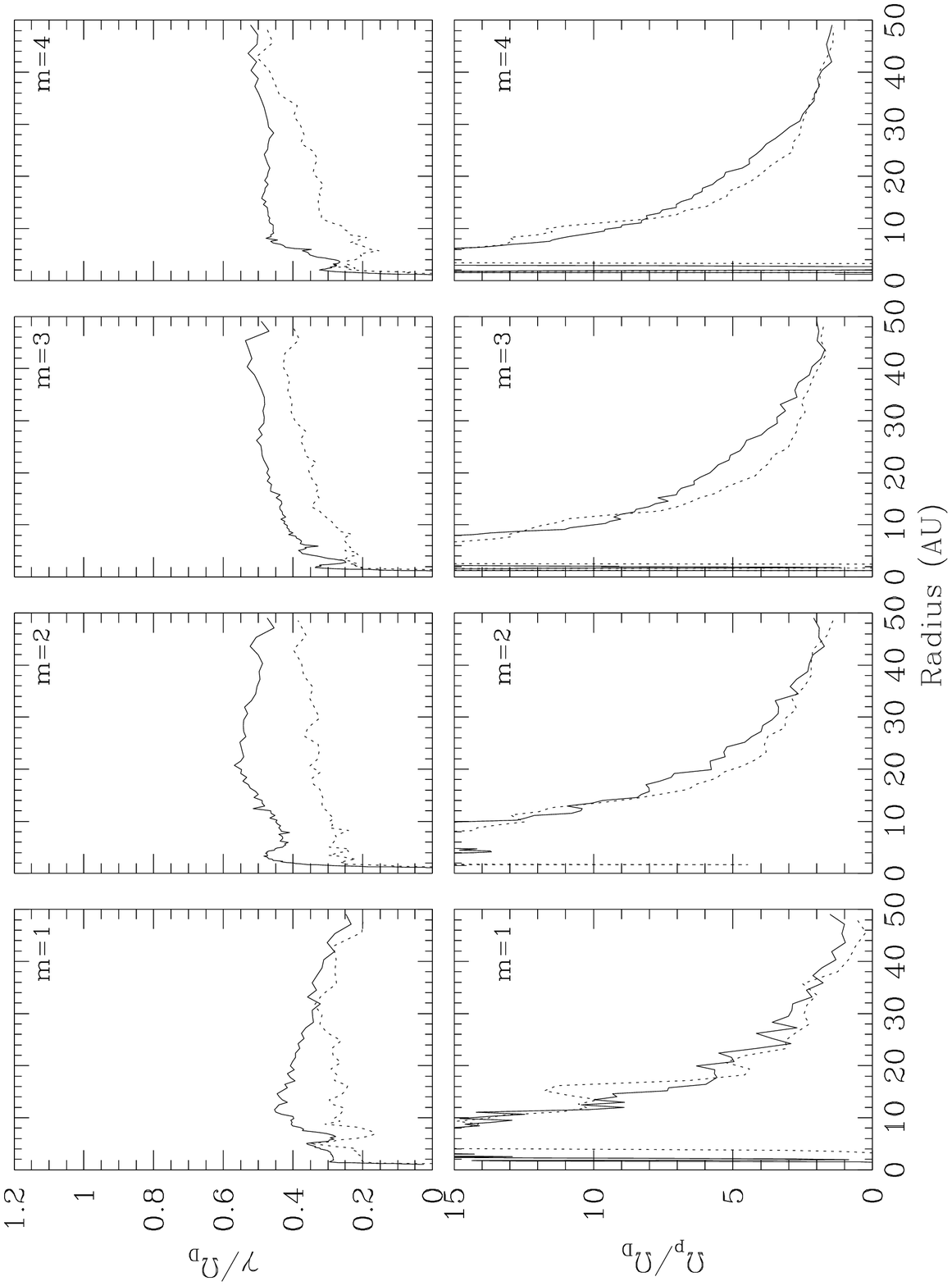}{5.5in}{-90}{70}{70}{-280}{500}
\caption{\label{patgrwm1-4low}
The growth rates and pattern speeds for the $m=1$--4 spiral
arm patterns. The simulation from which these are derived is the same
as is shown in figure \ref{ppm-lomas}. The solid lines represent the
moderate resolution simulation {\it pch2}  while the dotted lines 
represent results from the lower resolution simulation {\it pcm2}.  }
\end{figure}

\clearpage

\begin{figure}
\plotfiddle{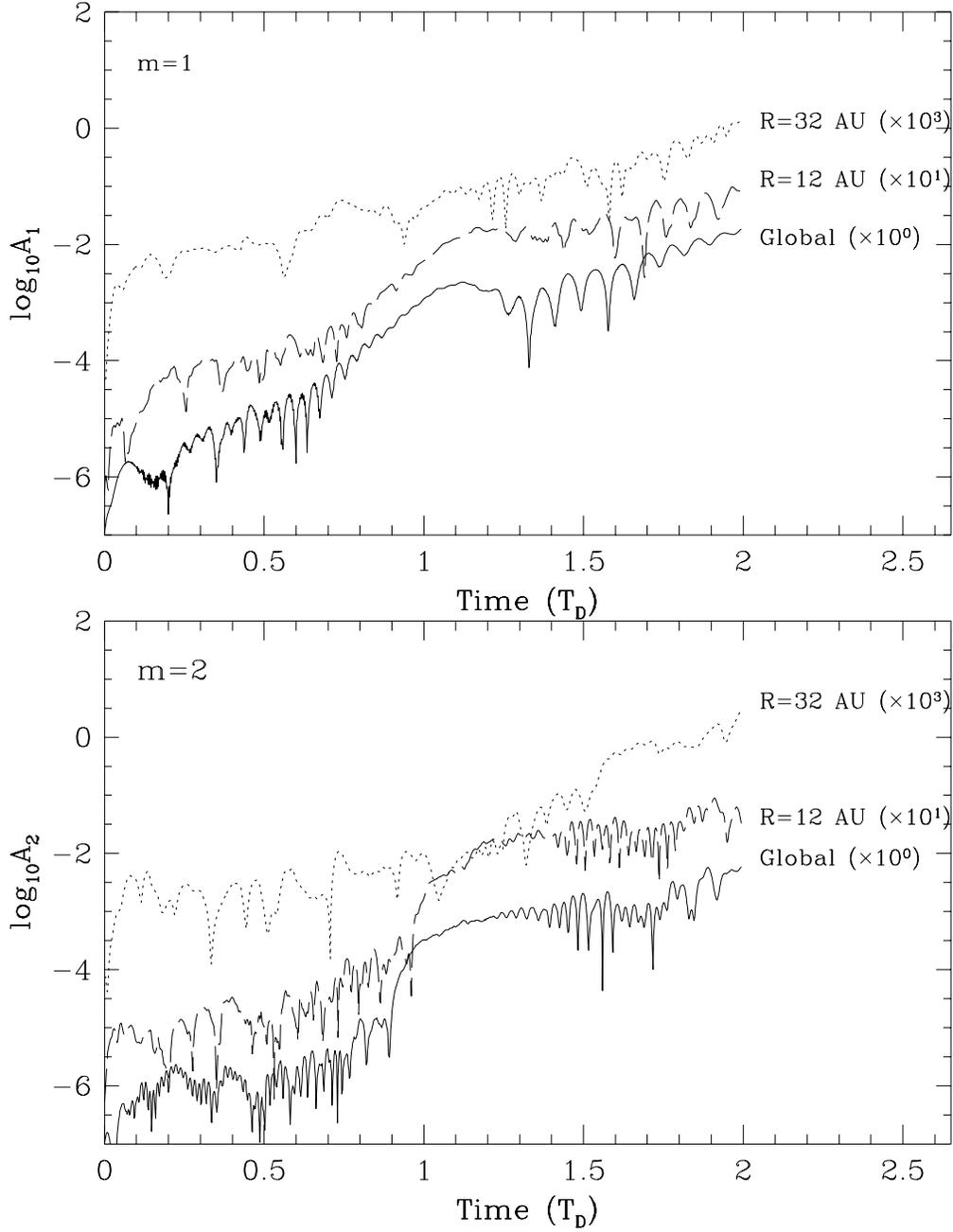}{6.0in}{0}{70}{70}{-220}{-30}
\caption{\label{hiq-amp}
Amplitude of the $m=1$ and $m=2$ spiral patterns at various locations in
the disk simulation {\it pqm5}. The outer portion of this disk is
initially quiescent.  The amplitude of the $m=1$ pattern does begin
growing immediately, however near $T_D\sim 1$ it experiences a `hump'
in its amplitude as instability propagates towards larger radii.
The region near the density maximum ($R\sim$12 AU) experiences little
initial growth in $m>1$ patterns, but once instabilities enter that
region (cf. the lower right panel of figure \ref{ppm-qvar}) they
quickly grow to dominate the instability amplitude over the entire
system. }
\end{figure}

\clearpage

\begin{figure}
\plotfiddle{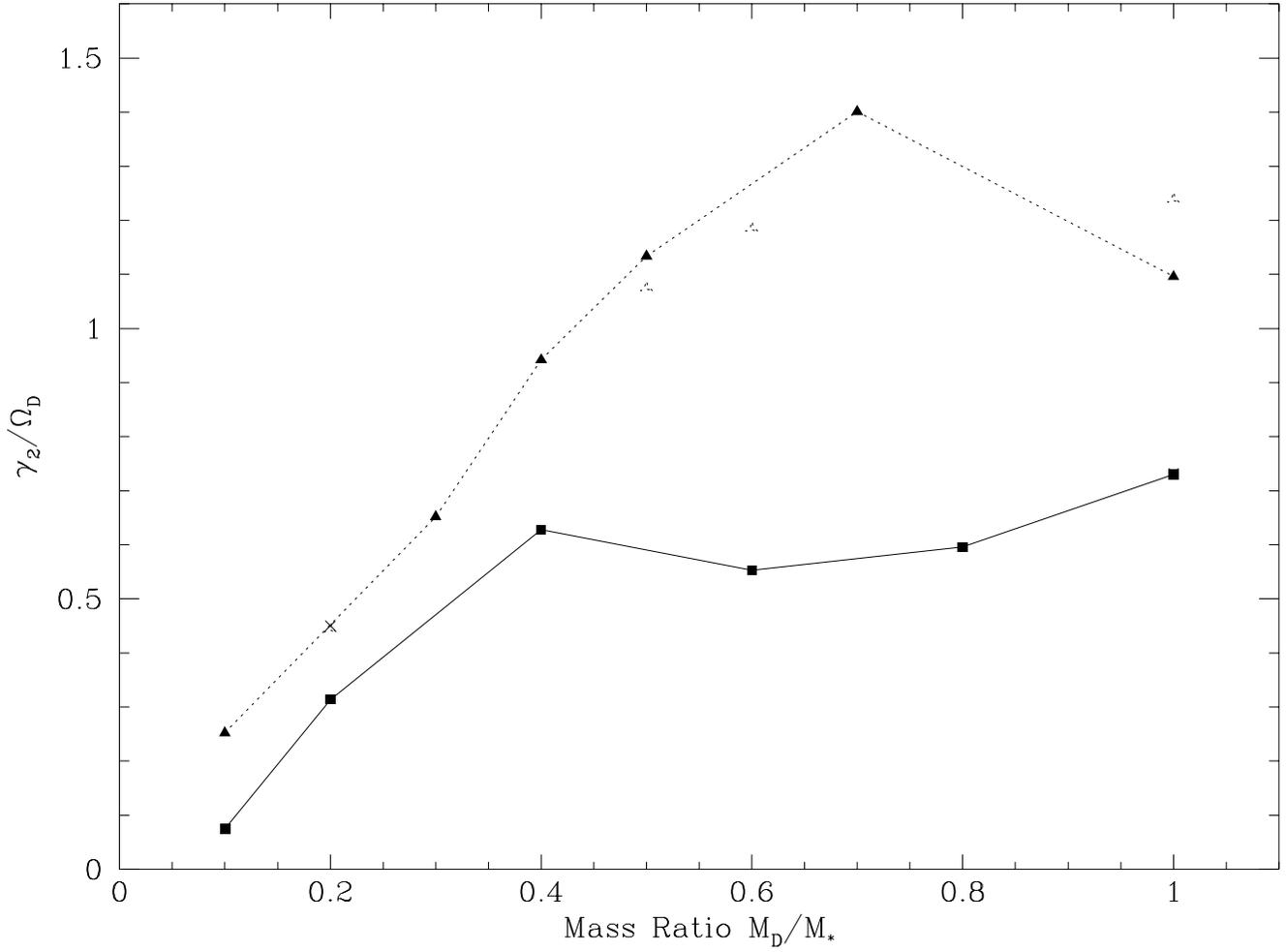}{5.5in}{-90}{70}{70}{-280}{440}
\caption{\label{disk-mratrates}
Growth rates for the m=2 mode for PPM simulations using a reflecting
outer boundary condition at moderate resolution (solid squares) and
at higher resolution ($\times$). A second series of simulations
with an infall boundary condition are shown with solid triangles
and at higher resolution with open triangles. }
\end{figure}

\clearpage

\begin{figure}
\plotfiddle{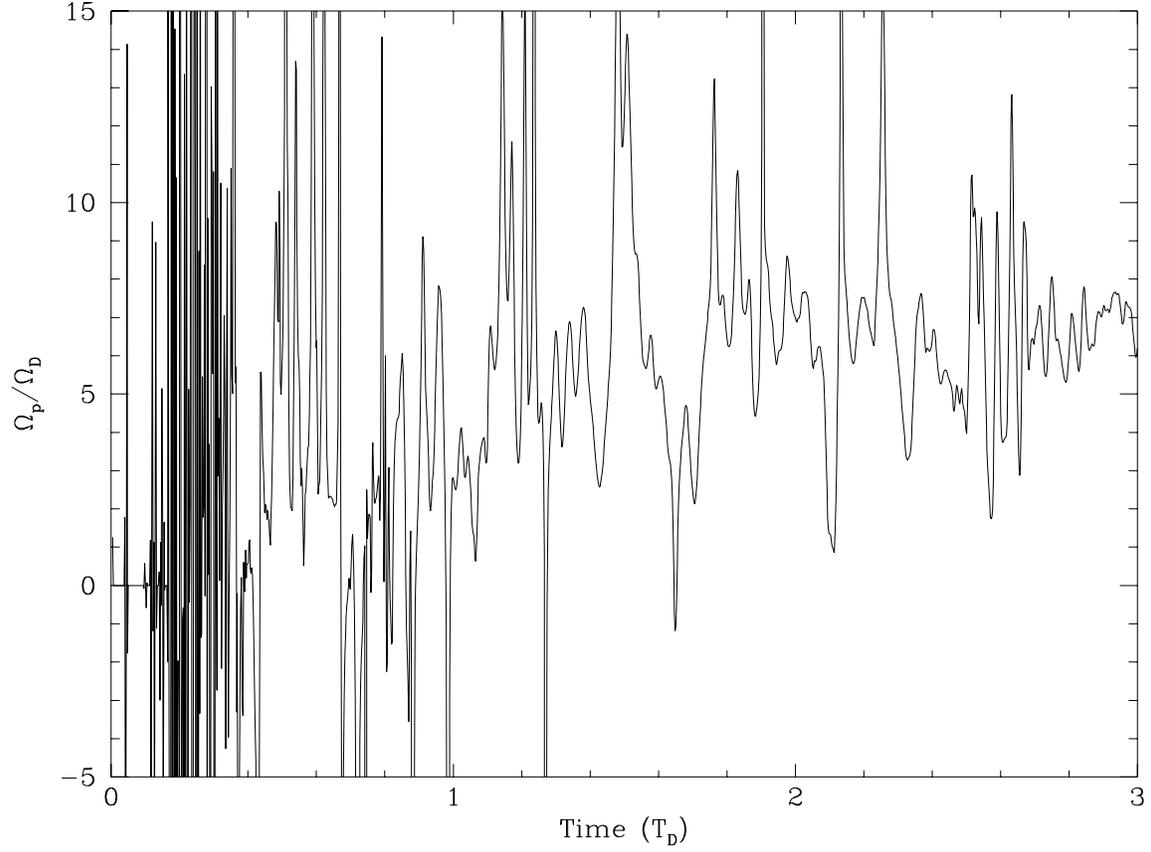}{5in}{-90}{60}{60}{-230}{400}
\caption{\label{m2pat_tme}
Pattern speeds for the $m=2$ pattern as a function of time for the
disk shown in figure \ref{ppm-himas}.  The pattern speed is for the
pattern at a radius $R\approx$32 AU from the star, which is near 
the middle of the region where the density is a power law in form. }
\end{figure}

\clearpage

\begin{figure}
\plotfiddle{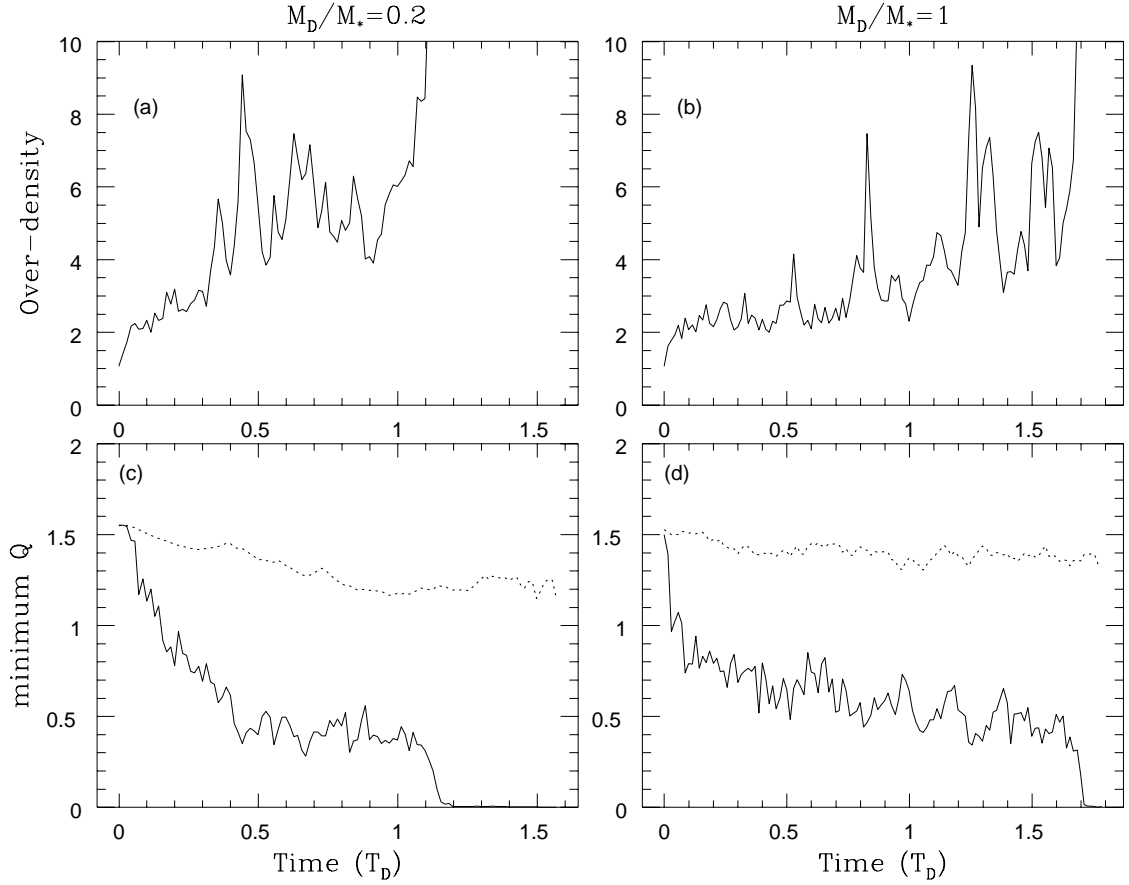}{5in}{-90}{60}{60}{-230}{400}
\caption{\label{odqplot}
Maximum over-density in SPH disks of low (a) and high (b)
disk/star mass ratio plotted vs. time (simulations {\it scv2} and
{\it scv6}).  Each disk begins with an initial \qmin~$=1.5$. 
Upon clumping the over-density assumes values 2-3 orders of 
magnitude larger than are plotted here and are omitted from these
graphs. (c) and (d) show the minimum $Q$ value for the same disks as
shown in (a) and (b) with both minimum azimuth average values 
(dotted line) and local minimum (solid line) values shown.}
\end{figure}

\clearpage

\begin{figure}
\plotfiddle{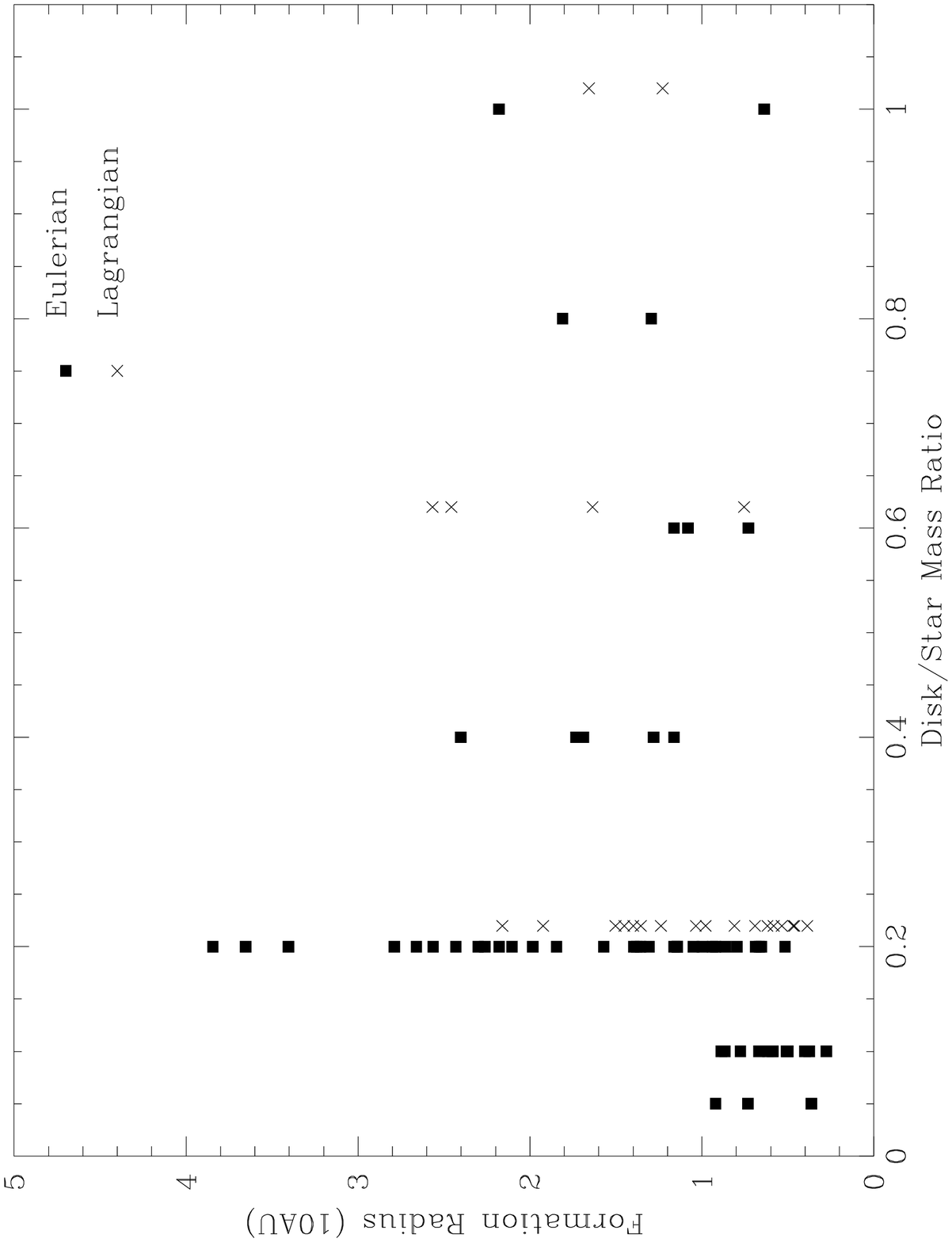}{5in}{-90}{60}{60}{-230}{400}
\caption{\label{formrad}
Formation radius (in units of 10 AU) for each clump vs. disk
mass. Each disk in the series {\it scv0-scv6} begins with an initial  
minimum $Q$ of 1.5.  Clumps form predominantly in the inner half of the 
disk, with only the \mrat~$=0.2$ disk showing clump formation over the
entire range in radius. In the simulations in which more than $\sim$10 
clumps formed an exact number becomes difficult to determine.  
Collisions between clumps and fission of a single clump into two (due 
to accretion of a large amount of angular momentum over a short 
time) make long term identification of any clump which has undergone 
a collision or fission event ambiguous and we do not include them
here. Filled triangles represent simulations evolved under an
Eulerian isothermal assumption (see section \ref{eos-sec}) while the
crosses (offset from their disk masses slightly to avoid confusion)
represent disks with the Lagrangian isothermal assumption. }
\end{figure}

%% file: disk.bbl
\newpage

\begin{references}

\reference{} Adams, F.C. \& Benz, W., 1992, Gravitational 
   Instabilities in Circumstellar Disks and the Formation of Binary 
   Companions, in Complementary Approaches to Double and Multiple 
   Star Research, IAU Colloquium No. 135, (Provo: Publications of the 
   Astr. Soc. of the Pac.)  (AB92)

\reference{} Adams, F. C., Emerson, J. P. \& Fuller, G. A. 1990, \apj, 
357, 606

\reference{} Adams, F. C. \& Lin, D. N. C., in Protostars and Planets
III, ed. Lunine, J. I. and Levy, E. H., Tucson: University of Arizona 
Press

\reference{} Adams, F. C., Ruden, S. P. \& Shu, F. H., 1989, \apj, 
347, 959 (ARS)

\reference{} Adams, F. C. \& Watkins, R., 1995, \apj, 451, 314

\reference{} Artymowicz, P., 1994, \apj, 423, 581

\reference{} Artymowicz, P., 1993, \apj, 419, 166

\reference{} Artymowicz, P., \& Lubow, S., 1994, \apj, 421, 651

\reference{} Balsara, D., 1995, J. Comp. Phys, 121, 357

\reference{} Bate, M., Bonnell, I. \& Price, N. M., 1996, \mnras, 
277, 362

\reference{} Beckwith S. V. W., Sargent, A. I., Chini, R. S. \&
G\"usten, R. 1990 \aj, 99, 924

\reference{} Benz, W., 1990, in  The Numerical Modeling of Nonlinear
   Stellar Pulsations p. 269, J. R. Buchler ed.

\reference{} Benz, W., Bowers, R. L., 
      Cameron, A. G. W. \& Press, W. H. 1990, \apj, 348, 647

\reference{} Binney, J., \& Tremaine, S., 1987, Galactic Dynamics,
      Princeton: Princeton University Press

\reference{} Boss, A., 1995, \apj, 439, 224

\reference{} Bonnell, I. \& Bastien, P., 1992, \apj, 401, 654

\reference{} Burkert, A. \& Bodenheimer, P., 1993, \mnras, 264, 798

\reference{} Cassen, P. M. \& Moosman A., 1981. Icarus, 48, 353

\reference{} Christodoulou, D. \& Narayan, R., 1992, \apj, 388, 451

\reference{} Colella P. \& Woodward, P. R., 1984, J. Comp. Phys., 54, 174

\reference{} Duquennoy, A., \& Mayor, M., \aap, 248, 485

\reference{} Foster, P. N. \& Boss, A. P., 1996, \apj, 468, 784

\reference{} Fryxell, B. A., M\"uller, E. \& Arnett W. D., 1989, Max Plank
Institut f\"ur Astrophysik Report \#449

\reference{} Fryxell, B. A., M\"uller, E. \& Arnett W. D., 1991, \apj, 
367, 619

\reference{} Gatewood, G., 1996, \baas, 28, 885

\reference{} Ghez, A., Neugebauer, G., \& Mathews K., 1993, \aj, 106,
2005

\reference{} Goldreich, P., \& Tremaine, S., 1980, \apj, 241, 425

\reference{} Heemskirk, M. H. M., Papaloizou, J. C. B. \& Savonjie, G. J.,
1992 \aap, 260, 161

\reference{} Herant, M., \& Woosley, S. E., 1994, \apj, 425, 814

\reference{} Laughlin, G. \& Bodenheimer, P., 1994 \apj, 436, 335

\reference{} Laughlin, G., Korchagin, V. \& Adams, F. C. 1996, \apj,
477, 410

\reference{} Laughlin, G., \& R\'o\.zyczka, M., 1996 \apj, 456, 279

\reference{} Leinert, Ch., Zinnecker, H., Weitzel, N., Christou, J.,
Ridgway, S. T., Jameson, R., Haas, M. \& Lenzen, R., 1993, \aap, 
278, 129

\reference{} Lynden-Bell D. \& Pringle, J. E., 1974, \mnras, 168, 603 
(LBP)

\reference{} Marcy, G. W., \& Butler, R. P., 1996, \apjl, 464, 147

\reference{} Mayor, M., \& Queloz, D., 1995, Nature, 378, 355

\reference{} Mayor, M., Queloz, D., Udry, S., Halbwachs, J.-L., 1997,
From Brown Dwarfs to Planets, in Astronomical and Biochemical Origins
and Search for Life in the Universe, IAU Colloquium No. 161, (Bologna,
Italy: Editrice Compositori).

\reference{} Monaghan, J. J., 1992, \araa, 30, 543

\reference{} Murray, J., 1994, PhD thesis: Monash University

\reference{} Myhill, E. A. \& Kaula, W. M., 1992, \apj, 386, 177

\reference{} Nakajima, T., Oppenheimer, B. R., Kulkarni, S. R.,
Golimowski, D. A., Matthews, K. \&  Durrance, S. T., 1995, Nature,
378, 463

\reference{} Ostriker, E. C., Shu, F. H., Adams, F. C., 1992, \apj 399, 192

\reference{} Papaloizou, J. C. B. \& Lin, D. N. C., 1989 \apj, 344, 645

\reference{} Papaloizou, J. C. B. \& Savonjie, G. J., 1991, \mnras, 248, 
353

\reference{} Pickett, B. K., Durisen, R. H., \& Davis, G. A., 1996, \apj,
458, 714

\reference{} Pickett, B. K., Cassen, P., Durisen, R. H., Link, R., 1998, \apj,
preprint

\reference{} Porter, D. H. \& Woodward, P. R., \apjs, 93, 309

\reference{} Shakura, N. J. \& Sunyaev, R. A., 1973, \aap,  24, 337

\reference{} Shu, F. H., Adams, F. C. \& Lizano, S., 1987 \araa,  25, 23

\reference{} Shu, F. H., Tremaine, S. Adams, F. C. \& Ruden, S. P., 1990 
              \apj,  358, 495  (STAR)

\reference{} Simon, M., Ghez, A. M., Leinert, Ch., Cassar, L., 
Chen, W. P., Howell, R. R., Jameson, R. F., Matthews, K., 
Neugebauer, G. \& Richichi, A., \apj,  443, 625

\reference{} Terebey, S., Shu, F. H. \& Cassen, P.,  1984 \apj,  286, 529.

\reference{} Woodward, J. W., Tohline, J. E. \& Hachisu, I., 1994, \apj,
 420, 247

\end{references}
